\documentclass[aps,prl,twocolumn,amsmath,showpacs,amssymb,superscriptaddress]{revtex4-1}

\usepackage[usenames, dvipsnames]{color}

\usepackage{graphicx}
\usepackage{epstopdf}
\usepackage{amsmath}
\usepackage{dcolumn}
\usepackage{soul}
\usepackage{chemformula}
\usepackage[colorlinks=true,linkcolor=blue,citecolor=blue]{hyperref}

\begin{document}

\title{Perspective: Atomistic Simulations of Water and Aqueous Systems\\ with Machine Learning Potentials}
\thanks{All authors contributed equally.}

\author{Amir Omranpour}
\affiliation{Lehrstuhl f\"ur Theoretische Chemie II, Ruhr-Universit\"at Bochum, 44780 Bochum, Germany}
\affiliation{Research Center Chemical Sciences and Sustainability, Research Alliance Ruhr, 44780 Bochum, Germany}
\author{Pablo Montero De Hijes}
\affiliation{University of Vienna, Faculty of Physics, Boltzmanngasse 5, A-1090 Vienna, Austria}
\affiliation{University of Vienna, Faculty of Earth Sciences, Geography and Astronomy,  Josef-Holaubuek-Platz 2, 1090 Vienna, Austria}
\author{J\"org Behler}
\email{joerg.behler@rub.de}
\affiliation{Lehrstuhl f\"ur Theoretische Chemie II, Ruhr-Universit\"at Bochum, 44780 Bochum, Germany}
\affiliation{Research Center Chemical Sciences and Sustainability, Research Alliance Ruhr, 44780 Bochum, Germany}
\author{Christoph Dellago}
\email{christoph.dellago@univie.ac.at}
\affiliation{University of Vienna, Faculty of Physics, Boltzmanngasse 5, A-1090 Vienna, Austria}
\thanks{All authors contributed equally.}
\date{\today}

\begin{abstract}
As the most important solvent, water has been at the center of interest since the advent of computer simulations. While early molecular dynamics and Monte Carlo simulations had to make use of simple model potentials to describe the atomic interactions, accurate ab initio molecular dynamics simulations relying on the first-principles calculation of the energies and forces have opened the way to predictive simulations of aqueous systems. Still, these simulations are very demanding, which prevents the study of complex systems and their properties. Modern machine learning potentials (MLPs) have now reached a mature state, allowing to overcome these limitations by combining the high accuracy of electronic structure calculations with the efficiency of empirical force fields. In this Perspective we  give a concise overview about the progress made in the simulation of water and aqueous systems employing MLPs, starting from early work on free molecules and clusters via bulk liquid water to electrolyte solutions and solid-liquid interfaces.
\end{abstract}


\maketitle

\section{Introduction}

A large fraction of the surface of the Earth is covered by water and, still, some ice, giving our planet its distinctive blue color when viewed from space. Water is carried down deep into the Earth's crust at subduction zones, influencing volcanism and plate tectonics, and in the atmosphere, in form of vapor, liquid or ice, water is a key climate factor from the troposphere up to the stratosphere and mesosphere. Down at the Earth's surface, water shapes landscapes, provides the basis for life and is central to many technologies that sustain humanity. Given its significance and abundance, it is no surprise that over the centuries much research has been undertaken to understand the properties of water and their physical origin. 

One of the central scientific questions addressed in water research is how the complex behavior of water, exhibiting many anomalies and a rich phase diagram, arises from the interactions of the chemically rather simple H$_2$O molecules. Due to the limited temporal and spatial resolution of many experimental probes, much of what we know about water has been learned from computer simulations. 
Specifically, atomistic simulations have provided detailed insights into the directed network of hydrogen bonds between molecules that governs the structure and dynamics of water and its interaction with solutes and surfaces \cite{P2723,P6630,P6629,P4571}. 
Moreover, computer simulations have made it possible to investigate water at extreme conditions that are not accessible in experiments. For instance, simulations have been used to study water and ice at pressure and temperature conditions prevailing in the deep Earth \cite{P6208} and in the interiors of the giant planets Uranus and Neptune  \cite{P0568}, as well as in the deeply supercooled state, the so-called ``no-man's land'', where crystallization occurs extremely quickly \cite{P4599,P6209}. 

Following the pioneering Monte Carlo (MC) simulations of Barker \cite{P4339} and molecular dynamics (MD) studies of Rahman and Stillinger \cite{P2723}  in the late 1960s and early 1970s, respectively, many computer simulations of water and aqueous systems were carried out. Initially, these simulations were based on empirical potentials \cite{P3629}, but later they relied increasingly on forces and energies obtained from electronic structure calculations \cite{P2088}. In the empirical potential -- or force field -- approach the functional form of the interaction potential is constructed to capture the main physical interactions between molecules, with parameters adapted to reproduce some experimentally known quantities and/or quantum mechanical reference data. 

Since the first water models for MD simulations were proposed more than half a century ago \cite{P6627,P6628,P2722},
a vast number of empirical potentials were developed for water and ice, ranging from simple forms based on pair interactions to sophisticated many-body potentials including polarization and charge transfer \cite{P6207,P2803,P2476,P3629,P6210,P4555,P6263,abascal2005potential}. Despite their often simple functional form, these models have been remarkably successful in capturing the key properties of water across the phase diagram~\cite{bore2022phase,montero2018viscosity}. Modelling chemical reactivity, however, has proven difficult using empirical potentials, and with a few exceptions \cite{P6206,P4378,P3906,EVB_Voth}, empirical water models are usually non-reactive. That is, they lack the capability to represent the dissociation and recombination of water molecules, which is, {\em e.g.}, essential for describing the famous Grotthuss mechanism of proton transport \cite{P1126}. The inclusion of proton transfer events is not only central for simulations of acids and bases but is also of utmost importance in countless chemical reactions with water involved as reactant, from hydrolysis of biomolecules to electrochemical water splitting. Nevertheless, to this day, empirical force fields have been used for the vast majority of simulations involving water, particularly in studies of biological macromolecules, were aqueous solvation effects are of crucial importance \cite{P6210}.

A more fundamental route to the computer simulation of water relies on determining interactions from first principles, {\em i.e.}, {\em ab initio} by solving the electronic Schrödinger equation. The first {\em ab initio} molecular dynamics simulations of liquid water were carried out 30 years ago based on density functional theory (DFT). Since then, {\em ab initio} methods have been used extensively to study water~\cite{P2088,P6211} and, more generally, aqueous systems \cite{P6212}. While in principle {\em ab initio} approaches have the potential for truly predictive simulations and also provide access to electronic properties, currently available approximate methods are still somewhat limited~\cite{P6270}. For instance, predicting the equilibrium density of liquid water has been proven difficult within DFT based on standard exchange correlation functionals, and only the inclusion of dispersion forces produces satisfactory results \cite{P6213,P6212}. Moreover, some attempts to study water using methods beyond DFT have been made, {\em e.g.} using MP2-based simulations~\cite{P3967} or employing the random phase approximation (RPA) \cite{P4504}.

Compared to empirical force fields, {\em ab initio} approaches are computationally more expensive by many orders of magnitude, severely limiting the accessible system size and simulation times. Hence, many processes of interest occurring in aqueous systems, for instance freezing, glassy dynamics, the solvation of complex interfaces and biopolymers, as well as complex chemical reactions, are far beyond the capabilities of current {\em ab initio} simulations. 
Therefore, in order to transfer the reliability of {\em ab initio} methods to more complex systems, in the past two decades a considerable effort has been made to develop efficient but accurate potential energy surfaces based on systematic and flexible functional forms~\cite{P2881,P4505,P6483,P6607}. In particular, the growing use of machine learning techniques such as artificial neural networks and Gaussian processes for the construction of highly accurate and efficient machine learning potentials (MLP) has revolutionized -- and continues to revolutionize -- the field of atomistic computer simulations~\cite{P2559,P3033,P4885,P6102,P6112,P6121,P5673,P5788,P5673,P5793,P4263,P6018,P5971,P6131,P5977,P6631}. These methodological advances in modern MLPs make it now possible to predict even complex properties of condensed systems from first principles, opening up exciting new possibilities in chemistry, materials science and related disciplines. 

\begin{figure}
\centering
\includegraphics[width=0.5\textwidth, trim= 20 0 20 0, clip=true]{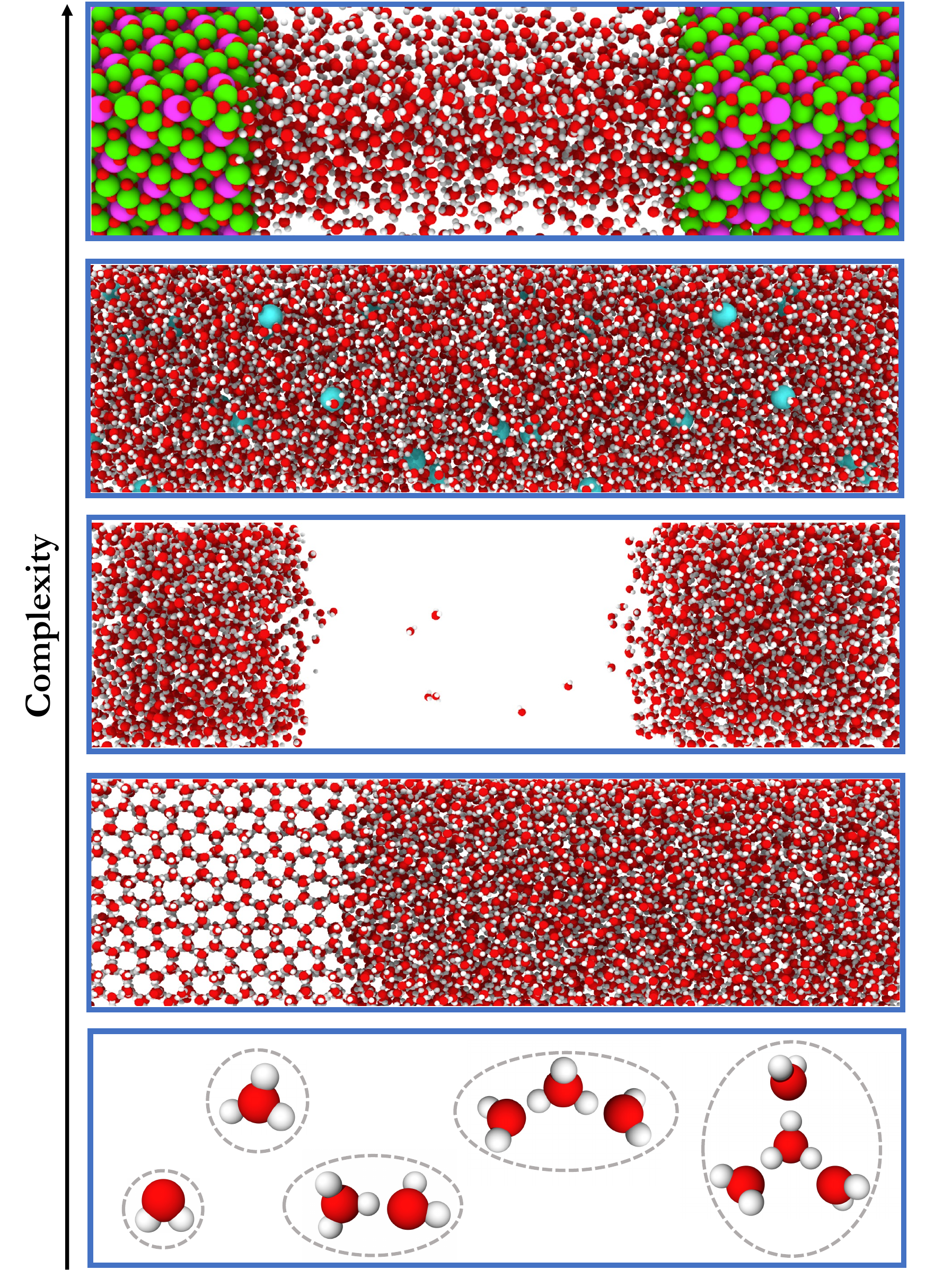}
\caption{Schematic overview of some important applications of MLPs to aqueous systems with increasing complexity (from bottom to top): neutral and protonated water clusters, liquid water and its interface with ice, liquid water–vapor interface, electrolyte solutions, and solid–water interfaces.}
\label{fig:complexity}
\end{figure}

Consequently, in recent years, atomistic modelling based on MLPs has also been increasingly applied to study water and, more generally, aqueous systems. The field has been evolving rapidly, both in terms of the underlying methodology and of the complexity of systems that can be addressed, from small water clusters in vacuum via bulk water and its phase diagram to solution chemistry and processes at solid-liquid interfaces (cf. Fig. \ref{fig:complexity}).

In this Perspective article we first provide a brief overview of the methodological basis and the current status of MLPs. In doing so, we focus on those types of MLPs that have been most frequently employed in simulations of systems involving water. In the subsequent sections, we then discuss different applications of MLPs to aqueous systems. While it is not the goal of this article to provide a comprehensive review covering all MLP-based studies of water and related systems exhaustively, we discuss a broad range of representative applications and point out future research directions to demonstrate the power and versatility of MLPs for the study of water and aqueous systems. 

%
\section{Methodology}

In recent years, machine learning potentials have become an increasingly important tool for atomistic simulations of complex systems in chemistry, physics and materials science. As a consequence, the development of MLPs is a very active topic of research and here we will restrict our discussion to a concise overview about the current status of the field. Readers interested in further details are referred to a very large number of reviews covering all aspects of the methodology of MLPs \cite{P2559,P3033,P4885,P6102,P6112,P6121,P5673,P5788,P5673,P5793,P4263,P6018,P5971,P6131,P5977,P6631,P6548}.

MLPs offer many advantages, like an excellent numerical agreement with the underlying electronic structure reference method, resulting in typical energy errors of only about 1~meV/atom and force errors in the order of 100~meV/\AA{}. These errors are significantly smaller than the uncertainty, {\em e.g.}, due to the choice of the exchange correlation functional in DFT, and thus replacing  electronic structure calculations by MLPs does only marginally affect the accuracy of the simulations. Moreover, MLPs can describe the making and breaking of chemical bonds, and provide a rather high computational efficiency enabling simulations of systems containing many thousands of atoms. Still, they are usually about one to two orders of magnitude slower to compute than simple classical force fields.

First MLPs have been introduced more than a quarter of a century ago by Doren and coworkers~\cite{P0316}, who suggested to use a feed-forward neural network (NN) to represent the interactions between diatomic molecules and solid surfaces. A general limitation of this first generation of MLPs, which has been further explored by numerous groups for about a decade for different types of systems, has been the limitation to a few atomic degrees of freedom only, restricting the applicability to small molecules in vacuum or small molecules interacting with frozen surfaces, as summarized in some early reviews~\cite{P2559,P3033}. The major challenge for extending this method to high-dimensional condensed systems like liquid water at that time has been the lack of suitable structural input descriptors for the machine learning algorithms, which ensure the imperative translational, rotational and permutational invariances of the potential energy surface (PES). Only for some applications system-specific approximate solutions could be derived when neglecting less important degrees of freedom of the system~\cite{P0830,P0421,P1388}. 
In parallel to these efforts in the development of early neural network-based MLPs, in pioneering work Braams and Bowman~\cite{P3062,P2881} introduced permutation invariant polynomials (PIP),  which are closely related and enable the construction of very accurate PESs by linear regression based on symmetrized polynomials as basis functions. While PIPs do not employ traditional machine learning algorithms like neural networks, they show a similar flexibility and include all invariances exactly, which has been a frustrating challenge for early MLPs employing non-linear models. Still, PIPs share with early MLPs the restriction to small systems with a very limited number of degrees of freedom. Over the years, PIPs have enabled the construction of very accurate potentials and have been applied successfully, e.g., to vibrations and reaction dynamics of small molecules in vacuum as well as a variety of water clusters~\cite{P2881}. 

MLPs have become generally applicable to high-dimensional systems containing thousands of atoms in 2007, when Behler and Parrinello introduced high-dimensional neural network potentials (HDNNP)~\cite{P1174,P5128,P4444,P4106}. The key step, which paved the way to study condensed systems like liquid water by MLPs, is the construction of the potential energy $E$ as a sum of environment-dependent local atomic energies $E_i$,
\begin{eqnarray}
E=\sum_{i=1}^{N_{\mathrm{atom}}}E_i, \label{eq:2GHDNNP}
\end{eqnarray}
where $N_{\rm atom}$ is the total number of atoms in the system. This general form of the total energy expression is shared by HDNNPs and many other second-generation MLPs proposed in the following years.

The energetically relevant local environment determining the $E_i$ is defined by a cutoff radius $R_{\mathrm{c}}$ such that all interactions beyond this radius, which is typically chosen between 5 and 10~\AA{}, are not explicitly included. While the ansatz of Eq.~\ref{eq:2GHDNNP} has been used in many empirical potentials for a long time, the introduction of second-generation MLPs has only become possible by the development of many-body descriptors with full translational, rotational and permutational invariance. In case of HDNNPs atom-centered symmetry functions (ACSF)~\cite{P2882} are most frequently used for this purpose, but nowadays a wide range of alternative descriptors is available and employed in different types of MLPs \cite{BartokPRB2013,PronobisJCTC2018,MusilChemRev2021,P5794}. All these descriptors represent structural fingerprints of the local atomic environments and serve as input for the machine learning algorithms, which then construct the functional relation between the atomic environments and the atomic energies. 

In case of HDNNPs, which are often used for simulations of water and aqueous systems, there is one feed-forward NN to be parameterized per element, which is then evaluated as many times as atoms of the respective element are present in the system. Closely related is ANI, which is a HDNNP with modified angular ACSFs~\cite{P4945} aiming for transferability across a wide range of organic molecules. Another NN-based MLP that is very frequently used for simulations of water is Deep Potential Molecular Dynamics (DeePMD)~\cite{P5596,P5367,P5076}, which employs a local atomic coordinate system and descriptors in this reference frame as input for the atomic NNs. Many other second-generation MLPs containing NNs have been proposed~\cite{P5792,P6605,P5792}, and in recent years also NN potentials learning descriptors as part of the training process employing message passing~\cite{P5368} have been put forward \cite{P4937,P5366,P5817,P6026,P6017,P6569,P6572}. Beyond neural networks, Gaussian approximation potentials (GAP) \cite{P2630} combined with the SOAP descriptor~\cite{P3885} are among the most frequently used MLPs, with a few applications to aqueous systems reported to date. Many other second-generation MLPs are available in the literature which can be expected to be used for systems containing water in the future~\cite{P4862,P4644,P5794}. 

An obvious limitation of second-generation MLPs is the truncation of the atomic interactions at the cutoff radius. However, in many aqueous systems, long-range electrostatic interactions play an important role~\cite{P5975}. These are explicitly considered in third-generation MLPs, which include electrostatic interactions employing environment-dependent charges represented by machine learning models. Already in 2007, Popelier and coworkers have shown that it is possible to construct electrostatic multipoles using neural networks~\cite{P2391} and Gaussian processes~\cite{P3199} to improve the description of electrostatics in classical force fields, and also applications to water clusters have been reported~\cite{P2392}.

In 2011, HDNNPs of the third-generation have been proposed by introducing a second set of atomic neural networks providing atomic partial charged trained to DFT reference data \cite{P2962,P3132}. From these charges the electrostatic energy can be computed and combined with the short-range expression of Eq.~\ref{eq:2GHDNNP} to yield the total energy of the system. By training the short-range part to represent only the energy component not covered by electrostatics, double counting of energy contributions can be avoided. Further MLPs including long-range electrostatics are, e.g., the HDNNP TensorMol~\cite{P5313}, the message passing network PhysNet~\cite{P5577}, and many others~\cite{P6056,P5885,P5372,P6200,P5205}. Nowadays, the machine learning representation of atomic partial charges and electrostatic multipoles is a very active field of research opening many routes to the construction of third-generation MLPs.

A remaining limitation of third-generation MLPs is the locality of the atomic charges, which does not allow to describe systems exhibiting long-range charge transfer and other non-local dependencies between the geometric and the electronic structure~\cite{P5977}. These phenomena can be considered in fourth-generation MLPs. The first MLP of this generation has been the charge equilibration neural network technique (CENT) proposed by Goedecker and coworkers in 2015~\cite{P4419}. Since the introduction of CENT, which employs a global charge equilibration step~\cite{P1448} and is intended for applications to ionic materials, several other fourth-generation MLPs have emerged, like Becke population neural networks (BpopNN)~\cite{P5859}, fourth-generation HDNNPs (4G-HDNNP)~\cite{P5932}, and charge recursive NNs (QRNN)~\cite{P6122}. To date, fourth-generation MLPs have not been extensively applied to water, but offer new interesting possibilities for studies of complex systems.

A key aspect in the training of MLPs~\cite{P6548} is the construction of suitable data sets covering the structures that are visited in the intended simulations, as MLPs often show a strongly reduced accuracy when extrapolating beyond the known part of configuration space. The size and composition of these data sets depend on the systems of interest, but due to the high flexibility of MLPs often energies and forces of 10,000 or more electronic structure calculations are required for training reliable potentials. For a systematic and unbiased determination of these structures, often various forms of active learning are employed~\cite{P5842,P5399,P5782,P6058}.

\section{Applications to Aqueous Systems} 
\subsection{Neutral and Protonated Water Clusters} 

Water clusters have received considerable attention already during the advent of MLPs as important benchmark systems. Even for small water clusters there is a large structural variety with many energetically close local minima, posing a significant challenge for potential development. At the same time, their moderate size allows to perform accurate high-level electronic structure reference calculations. 

Early first-generation MLPs for water clusters include an MP2-based six-dimensional PES for the water dimer with frozen monomer geometries reported in 1997 that made use of a single feed-forward NN~\cite{P0833}.  In 2006, a very accurate three-dimensional NN potential for the water monomer with an RMSE of only 1~cm$^{-1}$ (about 0.1 meV for the molecular potential energy) was published~\cite{P0826}, 
and also a NN potential for the same systems focusing on the permutation invariance of the PES has been reported in 2012~\cite{P3259}. From 2005 onwards, water clusters have also been investigated in great detail using permutation invariant polynomials reaching MP2 and coupled cluster  accuracies~\cite{P3140,P3139,P6161,P2296,P6160,P2716,P2720,P3958,P4365,P5641,P6321,P6645}.

To address larger water clusters, second-generation HDNNPs have been developed based on DFT data for a series of neutral clusters up to the decamer~\cite{P3875,P3971} as well as for several protonated clusters~\cite{P4472}. HDNNPs have also been applied to very large clusters containing hundreds of molecules~\cite{P6646}. Moreover, nuclear quantum effects in neutral and protonated water clusters were studied in recent years using very accurate HDNNPs trained to coupled cluster data~\cite{P5504,P5882,P5842,P6136,P6223,P5305}. 

Starting in 2009, NNs were employed using water clusters as test bed to find ways of improving the description of Coulomb interactions in classical force fields by learning environment-dependent electrostatic multipole moments~\cite{P2211}. A comparison of NNs and Gaussian process regression (GPR) for the representation of multipoles found these methods to be similar in terms of accuracy and costs~\cite{P2392}, resulting in an extension of the work to a water molecule embedded in a water decamer~\cite{P6632}. Finally, this approach has been developed further to the FFLUX water model and applied to larger water clusters~\cite{P6634}. In related work, GPR has been employed in a similar way to express charges in water molecules \cite{P5885}. 

A first MLP of the third generation expressing the full energy, {\em i.e.}, electrostatics and short-range bonding, of the system by machine learning has been a third-generation HDNNP for the water dimer reported in 2012~\cite{P3132}, which includes electrostatic interactions based on environment-dependent atomic partial charges expressed by a second set of atomic NNs.
Another example in which  electrostatics, as well as van der Waals interactions, were implemented on top of a short-range HDNNP for water clusters is TensorMol published in 2018~\cite{P5313}.
Moreover, small clusters have been used as benchmark for the inclusion of long-range interactions in DeePMD \cite{P6200}.

Finally, machine learning was used to learn tensorial properties using water clusters as benchmark~\cite{P5881,P6633}, and also the effect of noise on training HDNNPs for possible future applications on quantum computers has been tested for this system~\cite{P6635}. Beyond NNs and GPR, also Supports Vector Regression (SVR) has been explored for water clusters~\cite{P6636}. Although not directly used in the development of a high-dimensional PESs, the SVR method as well as Random Forest and Gaussian regression have also been used in specific applications like the prediction of the electron correlation energy in the water monomer and dimer \cite{P6637}. 
Beyond atomistic potentials, machine learning methods have also been used in various forms in combination with electronic structure concepts and methods like CCSD(T) and variational quantum Monte Carlo calculations with  benchmarks for monomers and small clusters~\cite{P6638,P5789,P6639,P6640,P6641,P6077,P6642}.

\subsection{Bulk Water and Ice}

Of all aqueous systems, the construction of MLPs for bulk liquid water has received most attention due to its crucial importance as solvent for a wide range of (bio)chemical reactions, but also because of its remarkable properties making water one of the most challenging benchmarks for the construction of interatomic potentials for molecular systems. 

Due to the limited dimensionality that could be dealt with at that time, the first application of machine learning for the simulation of pure liquid water in 2002~\cite{P0838} aimed for the inclusion of polarization in the TIP4P water model~\cite{P1472}. For this purpose a feed-forward neural network was used to represent the many-body interactions in dimers of rigid water molecules. This water model, named T4NN, was trained with MP2 reference data and was then used in Monte Carlo simulations to determine a range of properties of water such as its density, heat capacity and radial distribution functions. Overall, the agreement of many properties with experiment at standard conditions could be significantly improved with respect to the underlying TIP4P model while the transferability to different temperatures posed a challenge. 

In 2013, GAPs  
were used to enhance the accuracy of DFT-based {\em ab initio} MD simulations by introducing one- and two-body corrections trained on data of water monomers and dimers at the CCSD(T) level~\cite{P3872}. Their application to water clusters and MD simulations of liquid water showed an improved overall potential energy surface, but additional corrections beyond two-body terms were necessary for further improvements, as also demonstrated by comparisons to quantum Monte Carlo data~\cite{P3886}.

The first full-dimensional MLP for bulk liquid water and ice, which did not rely on a force-field or a DFT base-line potential, was reported in 2016~\cite{P4556}. In this work, a series of HDNNPs were trained using DFT reference data obtained for different GGA functionals, with and without dispersion corrections. This allowed, for the first time, to benchmark the quality of common DFT functionals in the description of computationally demanding properties of water such as the density anomaly, the melting temperature, the viscosity and the dielectric constant. The calculation of such quantities requires extensive simulations of large systems, which are prohibitively expensive using {\em ab initio} MD directly, but become affordable with HDNNPs. In this work, particular attention was devoted to study the effect of van der Waals interactions, which were shown to govern the flexibility of the hydrogen bond network and, hence, play a crucial role in determining the properties of water and ice. In fact, if van der Waals forces are neglected, the density maximum of water disappears and ice becomes denser than liquid water. In follow-up work, HDNNPs based on BLYP-D3 and RPBE-D3 data were used to study the density anomaly of water at negative pressures (for both functionals)~\cite{P5711} and the kinetics of the ice-water interface (only for RPBE-D3)~\cite{montero2023kinetics}.
  
Shortly after the first full-dimensional MLP for water, a HDNNP trained with B3LYP+D3 data was used in conjunction with path integral MD simulations to study nuclear quantum effects (NQE) of liquid water close to the triple point ~\cite{P4586}. Since then, several other studies of nuclear quantum fluctuations in bulk liquid water and ice have followed ~\cite{P4971,P6298,yao2020temperature,P6217}, including a thermodynamic stability analysis of liquid water as well as hexagonal and cubic ice employing hybrid DFT data~\cite{P5479}, and, more recently, a study of NQEs of liquid water based on the random phase approximation \cite{yao2021nuclear}. 

The past few years have witnessed a significant expansion in the use of MLPs for bulk water and ice. This growth encompasses many applications but also the development of methods, tools and extensive benchmarking. For instance, vibrational spectroscopy features of liquid water were extensively studied over the full frequency spectrum taking into account the effect of temperature and of overcoordinated hydrogen-bond environments employing a HDNNP based on revPBE-D3~\cite{P5570,P5639}. Embedded atom neural networks (EANN) have been used to represent tensorial properties in water describing vibrational features with the revPBE0-D3 functional~\cite{zhang2020efficient}. Moreover, using  water as a test example, DeePMD 
has been introduced in 2018~\cite{P5367}, which, alongside HDNNPs, emerged as one of the principal methods for modeling water using MLPs. DeePMD has also been coupled with empirical force fields~\cite{P6283}, and employed to develop coarse grained water models~\cite{P6280}. Some applications of DeePMD include the analysis of hydrogen bond dynamics in supercritical water~\cite{P6282}, the comparison of light and heavy water to assess isotope effects~\cite{P6268,P6291}, and the calculation of vibrational densities of states~\cite{liu2022toward}.
 
Crucial for the computation of vibrational features is the ability to determine the electronic polarizability tensor. In recent work, DeePMD was combined with an additional deep neural network to learn the environmental dependence of the polarizability tensor~\cite{P6196}. As demonstrated using the SCAN functional as reference, this approach yields accurate Raman spectra of liquid water. Furthermore, DeePMD has been integrated with a deep neural network trained to predict Wannier centers based on local environments. This approach allowed to compute infrared spectra~\cite{P6260,P6261} and to determine the temperature dependence of the dielectric constant~\cite{P6063}. The methods can be also extended to account for quadrupole moments~\cite{P6286}. Recently, the description of tensorial properties like the polarizability tensor has been fitted to MD simulations a posteriori using equivariant neural networks to describe infrared spectra~\cite{schienbein2023spectroscopy}. Moreover, a combination of HDNNP and GPR has allowed to model the hyper-Raman spectra of water, which helped to understand the differences in the OH stretch mode between infrared and Raman spectra~\cite{P6521}. 

The accuracy and computational efficiency of MLPs has made it possible to determine accurately thermodynamic properties of water including its phase diagram. For instance, the thermodynamic properties of water have been investigated with a DeePMD based on SCAN~\cite{P6278,P6151} and with a HDNNP based on revPBE0-D3~\cite{P5479,P6195}. The phase behavior of water under the extreme conditions expected in a planetary environment was also studied employing a HDNNP~\cite{P6264}. Other aspects addressed with MLPs include the study of heat transport~\cite{P6281,xu2023accurate} and the viscosity~\cite{P6285}. More recently, the thermodynamics of water has been investigated with a neuroevolution potential~\cite{chen2023thermodynamics}. 

Studying the mechanism and kinetics of phase transitions is computationally very demanding and thus completely out of reach for {\em  ab initio} simulations. MLPs, however, can be used to simulate systems of millions of water molecules with {\em ab initio} accuracy~\cite{P5960}, such that the simulation of phase transitions is now possible. One example is a recent investigation of the homogeneous nucleation of ice in supercooled water studied using DeePMD, in combination with the seeding methods \cite{bai2005test,sanz2013homogeneous}, in a system of hundreds of thousands of water molecules~\cite{P6284}. These calculations yielded nucleation rates consistent with experimental measurements. Very recently, advanced sampling techniques covering 36 $\mu s$ of total simulation time have been used to probe the atomic structure of the critical nucleus~\cite{chen2023imperfectly}. Also the liquid-liquid transition in supercooled water has been investigated. An initial attempt using a DeePMD based on the SCAN functional found indications of this transition through anomalies in thermodynamic response functions~\cite{P6273}. Two years later, the existence of this transition was conclusively demonstrated~\cite{gartner2022liquid} and its relation with the melting curves of ice polymorphs has been investigated~\cite{piaggi2023meltings}. Finally, building on previous works, the transition between ice Ih and its proton-ordered counterpart ice XI, mediated by ionic defects, has been studied based on the DeePMD model~\cite{P6266}. 

Since modern MLPs can capture reactions, important processes such as proton transfer and autoionization are accessible. In this regard, HDNNPs have allowed to describe
 the transport of hydronium and hydroxide ions including nuclear quantum effects~\cite{atsango2023developing} and the free energetics and mechanics of water dissociation~\cite{liu2023mechanistic},  allowing to compute the equilibrium $pK_w$ of  water \cite{calegari2023probing}.

The application of MLPs has been facilitated by careful benchmarking and transferability studies and the development of new ML-based methods and workflows. Some of these ML methods have been used beyond the fitting of potential energy surfaces as is the case for ML classifiers of phases~\cite{P3964,P5832,huang2022machine} and dynamical processes ~\cite{huang2022machine}. One particularly interesting finding regarding the transferability of MLPs is that liquid structures already contain the relevant information required to reproduce ice phases~\cite{P5968}, including even ice-water interfaces. In particular, this was observed in studies of homogeneous nucleation, in which an empirical potential was benchmarked against its MLP representation containing only liquid structures~\cite{guidarelli2023neural}. The performance of different density functionals (PBE, SCAN, vdW-cx, and optB88-vdW) for modeling water and ice has  been compared~\cite{P6265}. Moreover, HDNNP and GPR trained on the same dataset have been shown to be equivalent when compared over different thermodynamic properties of liquid water although HDNNPs seem to be more demanding in terms of the required training data~\cite{montero2023MLPs} but are computationally more efficient. In fact, the role of the training data has been the focus in other works~\cite{gomes2023size}. Furthermore, graph neural networks (GNN), which do not require predefined structural descriptors, have been applied to accelerate molecular dynamics simulations~\cite{li2022graph}. The  selection of descriptors  has also been automatized for HDNNPs~\cite{P5398,guidarelli2023finger}, whereas  GPR-based potentials have been employed in on-the-fly learning workflows~\cite{young2021transferable,montero2023MLPs}. 

Empirical potentials have also benefited from the development of MLPs. Coarse-grained MLPs  emerged~\cite{P6280,P5633,loeffler2020active,scherer2020kernel,thaler2021learning,musil2022quantum}, including an approach based on equivariant neural networks~\cite{loose2023coarse}, and empirical force fields were parameterized using ML algorithms~\cite{P6194,P6267,ye2021machine,wang2022machine}. Moreover, the addition of polarization to empirical force fields has been revisited, including charge transfer~\cite{han2023incorporating}. DeePMD has been used to fit an accurate but costly many-body potential \cite{P6492} reducing its computational cost by one order of magnitude~\cite{muniz2023neural}. Moreover, a GNN has been applied to estimate Bayesian uncertainty in molecular dynamics simulations based on an empirical potential~\cite{thaler2023scalable}. 

Further progress has come again from water clusters. HDNNPs, PIP-based potentials and GAPs have been shown to be equivalent in representing  many-body interactions in water clusters~\cite{P5360}, which can be employed in the construction of improved water potentials for bulk water. In fact, recent advances suggest that reference data obtained exclusively for water clusters could be sufficient to train accurate MLPs even for the bulk liquid phase~\cite{P5870,P6647,P6347,P6648,P6274}, including results from Gaussian-moment neural networks (GMNN)~\cite{P6274}. Recently, even gold-standard CCSD(T)-level accuracy for bulk water potentials has been reached by training to large clusters or periodic structures~\cite{P6360,P6321,P6362}.

An important current topic of research is the inclusion of long-range interactions \cite{P5975}, which are not explicitly considered in many MLPs. This problem has been addressed by introducing  non-local representations of the system remapped as local and equivariant feature vectors, capturing non-local and non-additive effects~\cite{grisafi2019incorporating}. Another approach to treat this issue is to learn the long-range response with a self-consistent field neural network, which has been shown to produce correct long-range polarization correlations in liquid water, as well as the correct response of liquid water to external electrostatic fields~\cite{P6272,dhattarwal2023dielectric}. 

Architectures like equivariant neural networks have been combined with empirical electrostatics and dispersion~\cite{ple2023force}. Such models are highly accurate in learning reference datasets~\cite{P6017,P6569,kovacs2023evaluation,kovacs2023mace,batatia2023foundation}, but their applicability has still to be tested in long simulations and large systems, and still some instability issues need to be solved~\cite{fu2022forces}. 

\subsection{Liquid/Vapor Interface}

Neural network potentials have also been used to investigate the structure, thermodynamics and spectroscopic properties of the liquid/vapor interface. As the local environments close to the interface are highly anisotropic and thus very different from the bulk, it is important that the training set explicitly includes data for interface configurations~\cite{P6150}. The structure of such configurations has been analyzed in detail using SOAP descriptors and local order parameters~\cite{donkor2023machine}.
Investigating the structure of the interface reveals the prevalence of orientations with the dipole moment roughly parallel to the surface with one OH bond pointing out of it~\cite{P6150}, corroborating insights gained from sum frequency generation (SFG) measurements~\cite{Bonn_Backus_2015}. 
By using the surface-sensitive velocity autocorrelation function~\cite{P6644}, such SFG spectra of the liquid/vapor interface were calculated from path integral molecular dynamics based on a HDNNP trained at the revPBE0+D3 level~\cite{P6134}. Recently, SFG spectra have been computed fully from first principles using a HDNNP combined with GPR~\cite{P6643}. 
 
In another study, it was found that a DeePMD potential relying only on local atomic energies can be applied to the liquid/vapor interface~\cite{P6152}. However, the explicit inclusion of long-range interactions was shown to be beneficial, confirming results of previous studies carried out for empirical potentials \cite{Sega2016}. The effect of long-range interactions was tested for a water molecule moving away from the liquid/vapor interface using an extension of DeePMD including long-range electrostatics~\cite{P6200}. The case of curved liquid/vapor interfaces has been addressed as well. For instance, DeePMD has been employed to investigate the formation of bubbles in metastable water~\cite{P6649}. Furthermore, it was shown that the free energy of water dissociation at the liquid/vapor interface of droplets and films deviates from the bulk, leading to an enrichment of hydronium cations at the interface and a depletion of hydroxide anions~\cite{P6650}.

\subsection{Electrolyte Solutions}

Beyond pure water, MLPs have been used in numerous simulations of electrolyte solutions~\cite{P6007}. Already in 1998, a feed-forward neural network was employed to represent the three-body interaction energies in \ch{H2O}--\ch{Al3+}--\ch{H2O} clusters with the aim to improve the force field description of \ch{Al3+} ions dissolved in bulk water~\cite{P0830}. This work represents an important milestone in the incorporation of permutation symmetry in structural descriptors. Later, HDNNPs have enabled the construction of full-dimensional DFT-quality PESs for aqueous NaOH solutions over the entire solubility range~\cite{P4670}. In this work, it has been found that as the NaOH concentration increases, the primary mechanism for proton transfer shifts from being acceptor-driven, influenced by the pre-solvation of hydroxide ions, to donor-driven, controlled by the pre-solvation of water molecules. Additionally, with increasing concentration, octahedral coordination geometries become less favored, in contrast to trigonal prism geometries \cite{P4895}. A novel water exchange mechanism has been identified around \ch{Na+}(aq) ions in basic (high pH) solutions \cite{P5126}. Studies comparing classical and ring-polymer molecular dynamics based on the HDNNP revealed that nuclear quantum effects significantly reduce proton transfer barriers, thus increasing proton transfer rates. This leads to an enhanced diffusion coefficient, especially for \ch{OH-}, and a shorter mean residence time of molecules in the first hydration shell around \ch{Na+} at high NaOH concentrations \cite{P5631}. Moreover, elevated temperatures in concentrated NaOH solutions amplify both the contributions of proton transfer to ionic conductivity and deviations from the Nernst–Einstein relation~\cite{P6138}. Further applications of HDNNPs include investigations of fluoride and sulfate ions in solution~\cite{P6058}. Employing similar methodologies and training on revPBE+D3 data, the dissolution mechanisms of NaCl in water have also been addressed~\cite{o2022crumbling}.
Another example is the use of HDNNPs to study zinc ion hydration in water~\cite{P5615}, with molecular dynamics simulations matching both the experimentally observed zinc-water radial distribution function and the X-ray absorption near edge structure spectrum. A genetic algorithm has also been utilized to study hydrated Zinc(II) ion clusters~\cite{wang2023structures}. In addition, microhydrated sodium ions with a few water molecules have been studied for both the potential energy and the dipole moment employing PIPs~\cite{P3582}.

DeePMD potentials have been used to study sodium chloride, potassium chloride, and sodium bromide at various concentrations~\cite{P6292}. These studies revealed that the structural changes due to the ions are confined to the immediate vicinity of the ions, where they disrupt the network of hydrogen bonds. Beyond these regions, the distribution of oxygen atoms relative to one another remains largely unchanged compared to pure water. In a related study, the dielectric permittivity of sodium chloride solutions has also been investigated~\cite{zhang2023dissolving}. Using DeePMD potentials, the uptake of \ch{N2O5} into aqueous aerosols has been examined, a process that is challenging to study experimentally due to the fast reaction kinetics of \ch{N2O5}~\cite{galib2021reactive}. Furthermore, the diffusivity of water in aqueous cesium iodide and sodium chloride solutions has been examined using a DeePMD framework trained on DFT data using the revPBE-D3 functional \cite{avula2023understanding}. Such simulations addressing the characteristic behavior of different ions are not readily accessible through traditional force field-based molecular dynamics simulations due to less ion-specific description of ion-water interactions.

\subsection{Water-Solid Interfaces}

Solid-liquid interfaces are of high interest for catalysis and electrochemistry. Due to the very different bonding in liquid water and in crystalline surfaces, such as metals or oxides, constructing unified atomistic potentials that can describe all subsystems of solid-liquid interfaces with balanced high accuracy presents a substantial challenge for empirical potentials. Moreover, in many cases, water is not only in contact with the surface but can also dissociate and recombine at a much higher rate than in the bulk liquid. Consequently, the use of reactive potentials, which can describe the making and breaking of bonds, is mandatory. MLPs are ideally suited for this purpose.

In 2014, an HDNNP for a thin water layer on top of 55-atom CuAu alloy clusters with varying stoichiometries and a slab model were reported to study the effect of water on the stability of different interface compositions by Monte Carlo simulations~\cite{P4028}. While this work exhibited a still rather large error of about 12 meV/atom, the accuracy of HDNNPs for solid-liquid interfaces has significantly improved in the following years. For instance, in studies of water at various surfaces of copper~\cite{P4859,P4886} and zinc oxide~\cite{P4988,P5601} energy RMSEs of less than 1~meV/atom could be reached. Detailed convergence tests with respect to the required system size were carried out~\cite{P4859}, showing that the diameter of the liquid phase between the slab surfaces needed to decouple the two surfaces by bulk-like water is at least about 35-40 \AA{}~\cite{P4859}. Such a size is beyond reach in \emph{ab initio} molecular dynamics but necessary to ensure that the central water molecules have bulk-like environments in their local vicinity. This consideration is crucial to ensure that these molecules do not experience any significant influence from the altered water structure near the interfaces or from the surfaces directly.

While it has been found using HDNNPs that water does not spontaneously react with defect-free surfaces of certain metals, such as copper, on nanosecond time scales~\cite{P4859}, fast dissociation and recombination processes leading to the formation of surface hydroxides have been observed on zinc oxide surfaces~\cite{P4988,P5601}. These processes are often governed by the surrounding hydrogen bond networks, significantly influencing the free energy barriers of proton transfer processes~\cite{P4988}. This phenomenon has been confirmed for water at TiO$_2$ surfaces using DeePMD potentials~\cite{P6271,P6473}, which have also been employed to explore the impact of slab thickness~\cite{P6481}. Furthermore, the use of HDNNPs for computing anharmonic frequencies has been suggested as a method to elucidate the role of hydrogen bonds in surface processes~\cite{P5600}.

Depending on the specific surface geometry, proton transfer events can lead to various topologies of proton transport networks along the surface, which can be either one-dimensional or two-dimensional. This has been demonstrated for several surfaces of zinc oxide~\cite{P5599} and the lithium intercalation compound LiMn$_2$O$_4$~\cite{P6141} using HDNNPs. Surface defects, often stabilized by solvation compared to the vacuum interface, have also been a subject of study. The mobility of adatoms has been found to significantly vary across different low-index surfaces of copper~\cite{P4886}. Investigations into the defective Zr$_7$O$_8$N$_4$/H$_2$O and pristine ZrO$_2$/H$_2$O interfaces using neural network potentials~\cite{nakanishi2023structural} revealed a bilayer water structure for Zr$_7$O$_8$N$_4$ and a monolayer structure for ZrO$_2$. Oxygen vacancies on the Zr$_7$O$_8$N$_4$ surface have been suggested as active sites for the oxygen reduction reaction. Furthermore, neural networks have been used to identify different oxidation states of transition metal ions at oxide-water interfaces~\cite{P6141}, which enables the characterization of electronic structures relevant for catalytic applications.

Due to its importance in catalysis, the TiO$_2$-water interface has so far been the most intensely studied interface using MLPs \cite{P6271,P6473,P6481,P6058,o2023elucidating,Li2023thermal,zeng2023mechanistic,ding2023modeling}. Investigations into the water coverage on the anatase (101) TiO$_2$ surface using the DeePMD potential \cite{o2023elucidating} have shown that higher water coverage prompts significant reorganization of the water monolayer at O$_2{}_c$ sites, leading to the formation of a two-dimensional hydrogen bond network with closely linked pairs of water molecules on neighboring TiO$_5{}_c$ and O$_2{}_c$ sites. Other DeePMD-based studies have examined the impact of water dissociation on thermal transport at the TiO$_2$–water interface~\cite{Li2023thermal}. While previous research on TiO$_2$-water interfaces mainly focused on the anatase (101) and rutile (110) TiO$_2$ surfaces, recent MLP-based studies~\cite{zeng2023mechanistic} have explored seven different TiO$_2$ surfaces using three distinct functionals: SCAN, PBE, and optB88-vdW. These studies found that water dissociation is more likely on the anatase (100), anatase (110), rutile (001), and rutile (011) surfaces, while molecular adsorption is the primary process on the anatase (101) and rutile (100) surfaces. Moreover, simulations for rutile (110) showed that the slab thickness significantly influences the results, with thicker slabs favoring molecular adsorption. DeePMD has also been used to study amorphous TiO$_2$ (a-TiO$_2$) to compare its behavior with well-studied crystalline TiO$_2$ at aqueous interfaces~\cite{ding2023modeling}. These studies demonstrated that water molecules on the a-TiO$_2$ surface do not exhibit the distinct layering typical of the aqueous interface of crystalline TiO$_2$. This difference results in an approximately tenfold increase in water diffusion speed at the interface.

Other cases of employing MLPs to study solid-water interfaces include the use of HDNNPs for the Pt(111)-water interface to investigate the interaction between water and hydroxylated metal surfaces~\cite{P6474,P6469}, and for the hematite-water interface~\cite{schienbein2022nanosecond}, revealing solvation dynamics at various time scales. DeePMD has been utilized for studying the TiS$_2$/water interface~\cite{li2023characterizing} to examine the influence of TiS$_2$ surface termination on the structure of interfacial water. Moreover, DeePMD has been applied to the IrO$_2$-water interface, exploring the hydration structure, proton transfer mechanisms, and acid-base characteristics~\cite{raman2023acid}, as well as to the GaP(110)-water interface~\cite{P6518}, which has been shown to require about 12 ns to reach equilibrium, a duration not achievable with traditional AIMD simulations. DeePMD has also been used for the construction of a potential aimed at studying ice nucleation at the microcline feldspar surface using the SCAN functional~\cite{piaggi2023first}, and for investigating the impact of water dissociation on thermal transport at the Cu-water interface~\cite{Li2023thermal.Cu}. Apart from HDNNPs and DeePMD, an equivariant graph neural network has been employed to study the oxygen reduction reaction at the Au(100)-water interface~\cite{P6470}, and on-the-fly learning kernel-based regression has been applied to investigate water adsorption on MgO and Fe$_3$O$_4$ surfaces, including surface reconstructions~\cite{li2020machine}.

Confined water, which exhibits properties notably differing from bulk water, has also been studied using MLPs. Some examples include HDNNPs for water confined between two-dimensional boron nitride sheets~\cite{P5876} and MoS$_2$\cite{P6058}, and water between graphite layers using DeePMD \cite{P6327,liu2023transferability} and committees of HDNNPs~\cite{kapil2022first}. Furthermore, MD simulations have been used to investigate water in single-walled carbon and boron nitride nanotubes~\cite{P6464,P6058}, finding a fivefold reduction in friction in carbon tubes compared to boron nitride, attributed to strong hydrogen-nitrogen interactions~\cite{P6464}. Ion concentration profiles under nanoconfinement~\cite{cao2023neural} have also been studied using neural network potentials, focusing on the effects of channel widths, ion molarity, and ion types.
   
\subsection{Other Systems}

Apart from studies of pure water, electrolytes and solid-liquid interfaces, which have been in the focus of MLP-based atomistic simulations for several years, the use of MLPs for aqueous systems has  increasingly diversified and now covers essentially all fields of simulations involving water. An exhaustive coverage of the related literature is beyond the scope of this Perspective, and in this section we just point the interested readers to several typical applications of MLPs in these rapidly growing fields.

A prominent use of atomistic simulations is to study chemical reactions of organic molecules in solution. 
Examples for the application of MLPs are corrections to QM/MM simulations of S$_N$2 reactions in water~\cite{P6145}, the solvation of protein fragments~\cite{P5577}, the decomposition of urea in water \cite{P6288}, the computation of free energy profiles of reactions of organic molecules~\cite{P6573}, and enzyme reactions~\cite{P6297}. Further studies include the quantum dynamics of an electron solvated in water~\cite{lan2021simulating} and the excited state of \ch{CH3NNCH3} surrounded by several water molecules~\cite{chen2019integrating}.

Still, the all-atom description of chemical processes in solution can be demanding and consequently simplified MLPs have been proposed as well. For example, the solvated alanine dipeptide~\cite{wang2019machine, husic2020coarse} and the folding/unfolding of chignolin~\cite{wang2019machine, krämer2023statistically} have been studied using the coarse-grained CGnet potential. Moreover, a DeepPot-SE model has been used to describe molecules under the influence of an implicit solvent~\cite{yao2023machine}. 

Additional applications of MLPs for aqueous systems include the study of diffusion in hydrogen hydrates \cite{P6133}, the determination of vibrational frequency shifts in formic acid C=O stretching and C=N stretching of MeCN in water \cite{yang2023machine}, retinoic acid in water \cite{boselt2021machine}, graph-convolutional neural networks for benchmarking sets of solutes and chemical reactions in water \cite{hofstetter2022graph}, hydration dynamics and IR spectroscopy of 4-fluorophenol \cite{salehi2022hydration}, zinc protein studies \cite{xu2021automatically}, conformational shifts of stacked heteroaromatics \cite{loeffler2021conformational}, and solvation free energy prediction of organic molecules in redox flow batteries \cite{gao2021graphical}.
 
\section{Discussion and Outlook}

\begin{figure}[tb!]
\centering
\includegraphics[width=8.5cm]{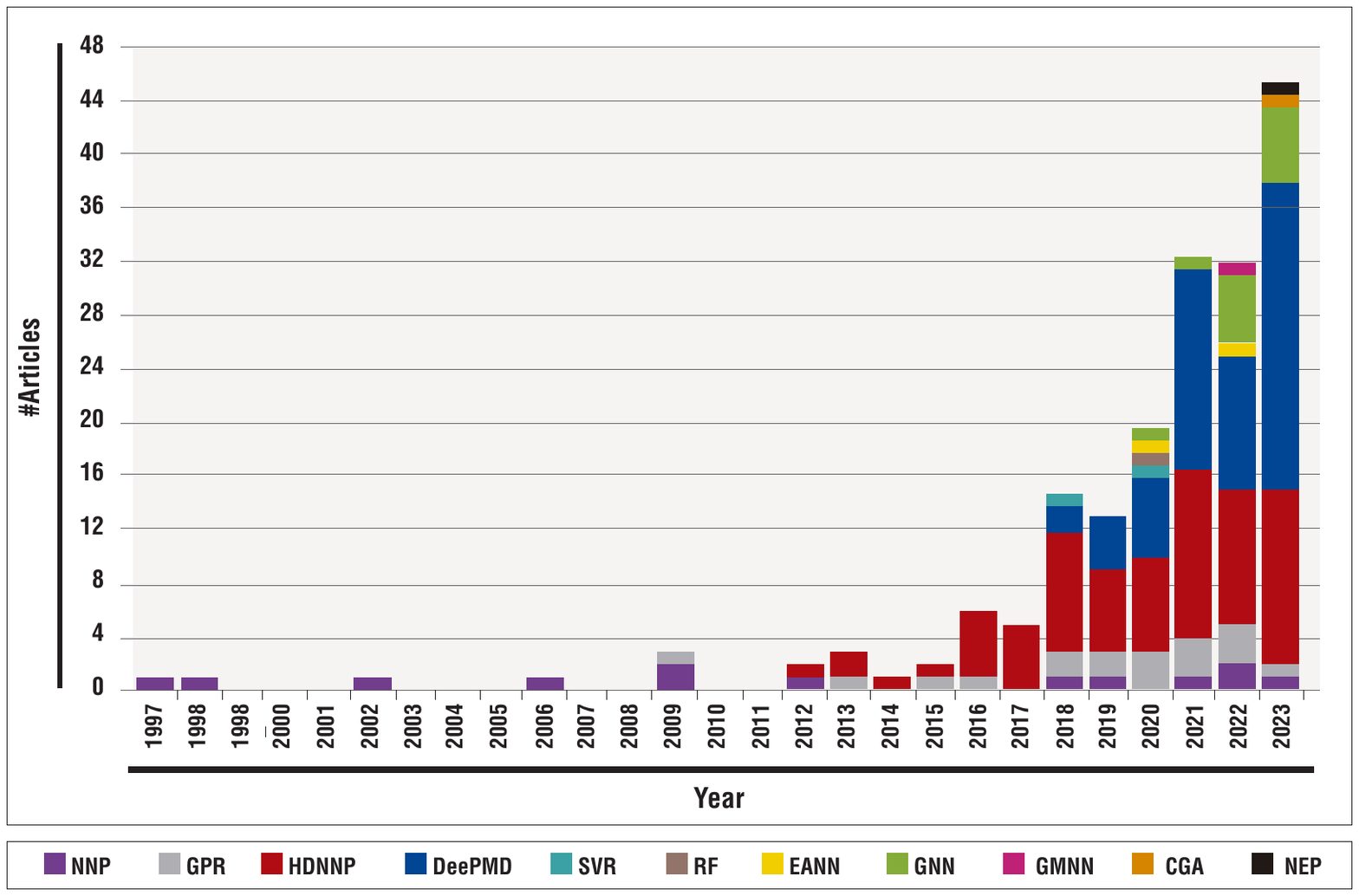}
\caption{ 
{Overview of the number of articles published per year for different types of MLPs applied to  water and aqueous systems as discussed in this Perspective. These are, in order of first use for these systems, neural network potentials (NNP) based on simple feed-forward neural networks, 
Gaussian process regression (GPR), high-dimensional neural network potentials (HDNNP), deep potential molecular dynamics (DeePMD), support vector regression (SVR),
random forest (RF), embedded atom neural networks (EANN), graph neural networks (GNN), Gaussian-moment neural networks (GMNN), comprehensive genetic algorithm (CGA), and neuroevolution potentials (NEP). 
 }}
\label{fig:histo}
\end{figure}

In recent years, MLPs have reached a high level of maturity and currently a transition from proof-of-concept and benchmark studies to practical simulations of a wide range of complex systems is taking place. Therefore, it can be anticipated that MLPs will allow to overcome the limitations of conventional methods like empirical potentials in terms of accuracy and \emph{ab initio} molecular dynamics in terms of efficiency paving the way to simulations of extended systems with unprecedented accuracy. 
MLPs have demonstrated this capability already for a broad spectrum of aqueous systems, ranging from neutral and protonated water clusters to bulk liquid water and ice, liquid/vapor interfaces and from electrolyte solutions to complex solid-water interfaces (see Fig. \ref{fig:complexity}). In all these studies, MLPs have enabled simulations with first principles accuracy that previously have been prohibitively demanding, as evidenced by a rapidly growing number of publications in the field shown in Fig. \ref{fig:histo}. 

Over the past two decades, progress in MLPs for aqueous systems has focused on different frontiers. While right from the start the highly flexible functional form of ML algorithms has enabled a numerically very accurate representation of the electronic structure reference data, a severe challenge in the early years has been the very limited number of degrees of freedom that could be considered. Only the development of modern descriptors for the atomic environments allowed to extend MLPs to condensed systems like liquid water with all their associated degrees of freedom. Recently, message passing neural networks~\cite{P5368} have become a promising alternative to the use of predefined descriptors opening many new exciting possibilities. Another frontier that has increasingly received attention in recent years is the incorporation of physical concepts into hitherto purely mathematical machine learning potentials, with many developments specifically aiming for improved descriptions of long-range electrostatic interactions, van der Waals forces, long-range charge transfer~\cite{P5977}, and electron densities~\cite{P6074}. This inclusion of physically meaningful terms, not only in the total energy expression but also in form of novel descriptors~\cite{P5629,P6542}, will further increase the accuracy and transferability of the potentials. 

To effectively simulate chemical processes in the aqueous phase, the underlying potential function must rely on accurate electronic structure calculations, since the chosen reference method represents a natural limit for the accuracy of MLPs. 
While coupled cluster accuracy has already been achieved in MLPs for liquid water~\cite{P6360,P6321,P6362}, which would have been unthinkable with conventional empirical potentials, reaching this gold-standard for more complex systems like solid-liquid interfaces is very challenging. Therefore, DFT will likely remain the dominant method for the reference electronic structure calculations of many systems in the foreseeable future, in particular for those systems involving solids. Although the generalized gradient approximation (GGA) has been the most commonly employed functional in the study of aqueous systems as they offer a good compromise between computational cost and accuracy, more and more powerful computing resources increasingly enable the use of more advanced and computationally expensive functionals, such as meta-GGA (particularly SCAN) and hybrid functionals, in the investigation of several aqueous systems, and even this level of accuracy has remained essentially inaccessible by on-the-fly \emph{ab initio} MD to date. Consequently, complex aqueous systems can now be investigated with a previously unattainable level of accuracy enabling predictive simulations. Moreover, MLPs can be employed to evaluate the accuracy of the underlying level of theory with respect to experimental values in a broad range of scenarios. This extends beyond the traditional comparison at standard temperature and pressure, allowing for a comprehensive evaluation that may provide insightful guidance for the development of theoretical methods. 

The functional flexibility is a key property of MLPs, but it is a double-edged sword: on the positive side, this flexibility enables the accurate approximation of the PES based on the reference data. On the negative side, however, such flexibility severely limits the extrapolation capabilities of MLPs to chemical environments not adequately sampled during the training process. In fact, extrapolation to unfamiliar environments can lead to unphysical structures and completely wrong simulation results. Therefore, the construction and validation of MLPs have to be done with great care to ensure that all  relevant local environments are included in the training set. Moreover, it is much more challenging than in case of simpler empirical potential to provide ``boxed'' MLPs for general usage, since not only the underlying parameters but also information about the range of validity is crucial information for successful applications. For instance, when studying solid-water interfaces, it becomes necessary to train the MLP not only on the bulk material and bulk water separately but also on systems that include all relevant interface configurations. Thus, increasing the complexity of the system also increases the number of local environments that must be included in the training data. Although AIMD simulations are sometimes used to generate initial training sets, they are usually insufficient since they often fail to capture the less frequently visited structures. Fortunately, this challenge has been largely overcome in recent years by incorporating active learning for the generation of the reference data. 


 While in this Perspective we have focused on the construction and application of accurate and efficient interatomic potentials, machine learning approaches can be useful in several other ways for the atomistic simulation of aqueous systems, and, more generally, of materials and biomolecular systems \cite{ARPC_Noe_2020}. For instance, neural networks have been employed to classify local structures in liquid water and various forms of ice \cite{JCP_Geiger_2013, JCIM_Fulford_2019}. The accurate identification of molecular structures with high spatial resolution is important, for instance,  in the study of crystallization, melting, crystal growth and the formation and migration of defect. Another challenge in molecular simulations lies in reducing or even completely removing correlations in the statistical sampling of configuration space as it occurs in sequential sampling methods such as molecular dynamics and Markov Chain Monte Carlo simulations. Here, normalizing flows \cite{noe2019boltzmann, MLST_Wirnsberger_2022} or other generative methods can play an important role for sampling the equilibrium distribution and also for creating new stable crystal structures~\cite{zeni2023mattergen}. Moreover, machine learning approaches have been suggested as a way to discover reaction coordinates and enhance the sampling of rare transitions occurring in complex molecular systems \cite{PNAS_Tiwary_2023, NatCompSci_Jung_2023}. There is little doubt that in the years to come further new ML and AI tools will be applied to the computational investigation of matter at the atomistic level, creating new opportunities for studying complex aqueous systems

%
\section{Conclusions}

Machine learning and, more generally, artificial intelligence is currently revolutionizing our way to do science and is providing new opportunities not achievable with traditional approaches. In the field of computational materials science, the advent of accurate, flexible and efficient MLPs has dramatically increased the time and length scale accessible by atomistic simulations of materials with {\em ab initio} accuracy. In this Perspective, we have provided an overview of the key concepts of such machine learned potentials with a focus on their application to water and aqueous systems. The broad spectrum of systems successfully explored with these techniques, including water clusters, bulk liquid water and ice, the liquid/vapor interface, electrolyte solutions, and solid-liquid interfaces, underscores the flexibility, efficiency and the high level of maturity they have reached since recent years. 

Since machined learned potentials accurately reproduce the underlying reference data obtained with electronic structure methods but in applications require a much lower computational effort, they now make it possible to compute complex materials properties such as phase diagrams. In this way, they do not only allow to gain new insights into a variety of systems, but they also provide a way to truly test the theoretical description underlying the reference data and to reveal their possible limitations. This facilitates the generation of high quality reference data in the future
as a basis for truly predictive ML-based computer simulations of complex materials. 

In summary, modern MLPs have created new opportunities for the investigation aqueous systems that would have been unimaginable with conventional methods for the foreseeable future. Carefully trained and validated MLPs can be employed to study complex reactive aqueous systems accurately across large time and length scales, without imposing {\em ad hoc} empirical constraints. While predicting the future of this rapidly evolving field is challenging, the remarkable progress made to date suggests that we can expect exciting new developments and some surprising breakthroughs in the years to come.

\section{Acknowledgements}
%
JB and AO are grateful for funding by the Deutsche Forschungsgemeinschaft (DFG, German Research Foundation) in TRR/CRC 247 (A10, project-ID 388390466) and under Germany's Excellence Strategy – EXC 2033 RESOLV (project-ID 390677874). CD and PMDH acknowledge funding from the Austrian Science Foundation FWF through the Projects Doctoral College Advanced Functional Materials (DOC 85-N) and the SFB TACO (F-81).
 
\section{Conflicts of interest}

The authors have no conflicts to disclose.

\section{Author Contributions}

{\bf Amir Omranpour:} Writing – original draft (equal); Writing – review \& editing (equal); Visualization (equal) {\bf Pablo Montero De Hijes:} Writing – original draft (equal); Writing – review \& editing (equal); Visualization (equal) {\bf J\"org Behler:} Conceptualization (equal); Writing – original draft (equal); Writing – review \& editing (equal); Resources  and Funding Acquisition (equal); Supervision (equal) {\bf Christoph Dellago:} Conceptualization (equal); Writing – original draft (equal); Writing – review \& editing (equal); Resources and Funding Acquisition (equal); Supervision (equal)

\bibliography{main} 

\begin{thebibliography}{100}

\bibitem{P2723}
A.~Rahman and F.~H. Stillinger, ``Molecular dynamics study of liquid water,''
  {\em J. Chem. Phys.}, vol.~55, no.~7, pp.~3336--3359, 1971.

\bibitem{P6630}
A.~Luzar and D.~Chandler, ``Hydrogen-bond kinetics in liquid water,'' {\em
  Nature}, vol.~379, no.~6560, pp.~55--57, 1996.

\bibitem{P6629}
K.~A. Dill, T.~M. Truskett, V.~Vlachy, and B.~Hribar-Lee, ``Modeling water, the
  hydrophobic effect, and ion solvation,'' {\em Annu. Rev. Biophys. Biomol.
  Struct.}, vol.~34, no.~1, pp.~173--199, 2005.
\newblock PMID: 15869376.

\bibitem{P4571}
O.~Bj{ö}rneholm, M.~H. Hansen, A.~Hodgson, L.-M. Liu, D.~T. Limmer,
  A.~Michaelides, P.~Pedevilla, J.~Rossmeisl, H.~Shen, G.~Tocci, E.~Tyrode,
  M.-M. Walz, J.~Werner, and H.~Bluhm, ``Water at interfaces,'' {\em Chem.
  Rev.}, vol.~116, pp.~7698--7726, 07 2016.

\bibitem{P6208}
V.~Rozsa, D.~Pan, F.~Giberti, and G.~Galli, ``Ab initio spectroscopy and ionic
  conductivity of water under earth mantle conditions,'' {\em Proc. Natl. Acad.
  Sci. U. S. A.}, vol.~115, no.~27, pp.~6952--6957, 2018.

\bibitem{P0568}
C.~Cavazzoni, G.~L. Chiarotti, S.~Scandolo, E.~Tosatti, M.~Bernasconi, and
  M.~Parrinello, ``Superionic and metallic states of water and ammonia at giant
  planet conditions,'' {\em Science}, vol.~283, no.~5398, pp.~44--46, 1999.

\bibitem{P4599}
P.~Gallo, K.~Amann-Winkel, C.~A. Angell, M.~A. Anisimov, F.~Caupin,
  C.~Chakravarty, E.~Lascaris, T.~Loerting, A.~Z. Panagiotopoulos, J.~Russo,
  J.~A. Sellberg, H.~E. Stanley, H.~Tanaka, C.~Vega, L.~Xu, and L.~G.~M.
  Pettersson, ``Water: A tale of two liquids,'' {\em Chem. Rev.}, vol.~116,
  no.~13, pp.~7463--7500, 2016.
\newblock PMID: 27380438.

\bibitem{P6209}
P.~G. Debenedetti, F.~Sciortino, and G.~H. Zerze, ``Second critical point in
  two realistic models of water,'' {\em Science}, vol.~369, no.~6501,
  pp.~289--292, 2020.

\bibitem{P4339}
J.~Barker and R.~Watts, ``Structure of water; a monte carlo calculation,'' {\em
  Chem. Phys. Lett.}, vol.~3, no.~3, pp.~144--145, 1969.

\bibitem{P3629}
C.~Vega and J.~L.~F. Abascal, ``Simulating water with rigid non-polarizable
  models: a general perspective,'' {\em Phys. Chem. Chem. Phys.}, vol.~13,
  pp.~19663--19688, 2011.

\bibitem{P2088}
K.~Laasonen, M.~Sprik, M.~Parrinello, and R.~Car, ``‘‘ab initio’’
  liquid water,'' {\em J. Chem. Phys.}, vol.~99, no.~11, pp.~9080--9089, 1993.

\bibitem{P6627}
J.~S. Rowlinson, ``The lattice energy of ice and the second virial coefficient
  of water vapour,'' {\em Trans. Faraday Soc.}, vol.~47, pp.~120--129, 1951.

\bibitem{P6628}
A.~Ben-Naim and F.~Stillinger, ``Aspects of the statistical-mechanical theory
  of water,'' in {\em Structure and transport processes in water and aqueous
  solutions} (R.~A. Horne, ed.), Wiley-Interscience, New York, 1972.

\bibitem{P2722}
F.~H. Stillinger and A.~Rahman, ``Improved simulation of liquid water by
  molecular dynamics,'' {\em J. Chem. Phys.}, vol.~60, no.~4, pp.~1545--1557,
  1974.

\bibitem{P6207}
F.~H. Stillinger, {\em Theory and Molecular Models for Water}, pp.~1--101.
\newblock John Wiley \& Sons, Ltd, 1975.

\bibitem{P2803}
B.~Guillot, ``A reappraisal of what we have learnt during three decades of
  computer simulations on water,'' {\em J. Mol. Liq.}, vol.~101, no.~1,
  pp.~219--260, 2002.
\newblock Molecular Liquids. Water at the New Millenium.

\bibitem{P2476}
C.~Vega, J.~L.~F. Abascal, M.~M. Conde, and J.~L. Aragones, ``What ice can
  teach us about water interactions: a critical comparison of the performance
  of different water models,'' {\em Faraday Discuss.}, vol.~141, pp.~251--276,
  2009.

\bibitem{P6210}
A.~V. Onufriev and S.~Izadi, ``Water models for biomolecular simulations,''
  {\em WIREs-Comput. Mol. Sci.}, vol.~8, no.~2, p.~e1347, 2018.

\bibitem{P4555}
G.~A. Cisneros, K.~T. Wikfeldt, L.~Ojamäe, J.~Lu, Y.~Xu, H.~Torabifard, A.~P.
  Bartók, G.~Csányi, V.~Molinero, and F.~Paesani, ``Modeling molecular
  interactions in water: From pairwise to many-body potential energy
  functions,'' {\em Chem. Rev.}, vol.~116, no.~13, pp.~7501--7528, 2016.

\bibitem{P6263}
T.~E. Gartner~III, K.~M. Hunter, E.~Lambros, A.~Caruso, M.~Riera, G.~R.
  Medders, A.~Z. Panagiotopoulos, P.~G. Debenedetti, and F.~Paesani,
  ``Anomalies and local structure of liquid water from boiling to the
  supercooled regime as predicted by the many-body mb-pol model,'' {\em J.
  Phys. Chem. Lett.}, vol.~13, no.~16, pp.~3652--3658, 2022.

\bibitem{abascal2005potential}
J.~Abascal, E.~Sanz, R.~Garc{\'\i}a~Fern{\'a}ndez, and C.~Vega, ``A potential
  model for the study of ices and amorphous water: Tip4p/ice,'' {\em J. Chem.
  Phys.}, vol.~122, no.~23, 2005.

\bibitem{bore2022phase}
S.~L. Bore, P.~M. Piaggi, R.~Car, and F.~Paesani, ``Phase diagram of the
  tip4p/ice water model by enhanced sampling simulations,'' {\em J. Chem.
  Phys.}, vol.~157, no.~5, 2022.

\bibitem{montero2018viscosity}
P.~Montero~de Hijes, E.~Sanz, L.~Joly, C.~Valeriani, and F.~Caupin, ``Viscosity
  and self-diffusion of supercooled and stretched water from molecular dynamics
  simulations,'' {\em J. Chem. Phys.}, vol.~149, no.~9, 2018.

\bibitem{P6206}
F.~H. Stillinger and C.~W. David, ``Polarization model for water and its ionic
  dissociation products,'' {\em J. Chem. Phys.}, vol.~69, no.~4,
  pp.~1473--1484, 1978.

\bibitem{P4378}
L.~Ojamäe, I.~Shavitt, and S.~J. Singer, ``Potential models for simulations of
  the solvated proton in water,'' {\em J. Chem. Phys.}, vol.~109, no.~13,
  pp.~5547--5564, 1998.

\bibitem{P3906}
M.~Raju, S.-Y. Kim, A.~C.~T. van Duin, and K.~A. Fichthorn, ``Reaxff reactive
  force field study of the dissociation of water on titania surfaces,'' {\em J.
  Phys. Chem. C}, vol.~117, p.~10558, 2013.

\bibitem{EVB_Voth}
U.~W. Schmitt and G.~A. Voth, ``Multistate empirical valence bond model for
  proton transport in water,'' {\em J. Phys. Chem. B}, vol.~102,
  pp.~5547--5551, 07 1998.

\bibitem{P1126}
D.~Marx, ``Proton transfer 200 years after von grotthuss: Insights from ab
  initio simulations,'' {\em ChemPhysChem}, vol.~7, pp.~1848--1870, 2006.

\bibitem{P6211}
M.~E. Tuckerman, K.~Laasonen, M.~Sprik, and M.~Parrinello, ``Ab initio
  simulations of water and water ions,'' {\em J. Phys. Condens. Matter},
  vol.~6, pp.~A93--A100, jun 1994.

\bibitem{P6212}
M.~J. Gillan, D.~Alfè, and A.~Michaelides, ``Perspective: How good is dft for
  water?,'' {\em J. Chem. Phys.}, vol.~144, no.~13, p.~130901, 2016.

\bibitem{P6270}
J.~G. Brandenburg, A.~Zen, D.~Alfè, and A.~Michaelides, ``Interaction between
  water and carbon nanostructures: How good are current density functional
  approximations?,'' {\em J. Chem. Phys.}, vol.~151, no.~16, p.~164702, 2019.

\bibitem{P6213}
A.~P. Gaiduk, F.~Gygi, and G.~Galli, ``Density and compressibility of liquid
  water and ice from first-principles simulations with hybrid functionals,''
  {\em J. Phys. Chem. Lett.}, vol.~6, no.~15, pp.~2902--2908, 2015.

\bibitem{P3967}
M.~D. Ben, M.~Schönherr, J.~Hutter, and J.~VandeVondele, ``Bulk liquid water
  at ambient temperature and pressure from mp2 theory,'' {\em J. Phys. Chem.
  Lett.}, vol.~4, p.~3753, 2013.

\bibitem{P4504}
M.~D. Ben, J.~Hutter, and J.~VandeVondele, ``Probing the structural and
  dynamical properties of liquid water with models including non-local electron
  correlation,'' {\em J. Chem. Phys.}, vol.~143, p.~054506, 2015.

\bibitem{P2881}
B.~J. Braams and J.~M. Bowman, ``Permutationally invariant potential energy
  surfaces in high dimensionality,'' {\em Int. Rev. Phys. Chem.}, vol.~28,
  pp.~577--606, 2009.

\bibitem{P4505}
G.~R. Medders, V.~Babin, and F.~Paesani, ``Development of a
  “first-principles” water potential with flexible monomers. iii. liquid
  phase properties,'' {\em J. Chem. Theory Comput.}, vol.~10, no.~8,
  pp.~2906--2910, 2014.

\bibitem{P6483}
E.~Palos, S.~Dasgupta, E.~Lambros, and F.~Paesani, ``Data-driven many-body
  potentials from density functional theory for aqueous phase chemistry,'' {\em
  Chem. Phys. Rev.}, vol.~4, p.~011301, 2023.

\bibitem{P6607}
Q.~Yu, C.~Qu, P.~L. Houston, A.~Nandi, P.~Pandey, R.~Conte, and J.~M. Bowman,
  ``A status report on gold standard machine-learned potentials for water,''
  {\em J. Phys. Chem. Lett.}, vol.~14, pp.~8077--8087, 2023.

\bibitem{P2559}
C.~M. Handley and P.~L.~A. Popelier, ``Potential energy surfaces fitted by
  artificial neural networks,'' {\em J. Phys. Chem. A}, vol.~114,
  pp.~3371--3383, 2010.

\bibitem{P3033}
J.~Behler, ``Neural network potential-energy surfaces in chemistry: a tool for
  large-scale simulations,'' {\em Phys. Chem. Chem. Phys.}, vol.~13,
  pp.~17930--17955, 2011.

\bibitem{P4885}
J.~Behler, ``Perspective: Machine learning potentials for atomistic
  simulations,'' {\em J. Chem. Phys.}, vol.~145, p.~170901, 2016.

\bibitem{P6102}
O.~T. Unke, S.~Chmiela, H.~E. Sauceda, M.~Gastegger, I.~Poltavsky, K.~T.
  Schütt, A.~Tkatchenko, and K.-R. Müller, ``Machine learning force fields,''
  {\em Chem. Rev.}, vol.~121, no.~16, pp.~10142--10186, 2021.
\newblock PMID: 33705118.

\bibitem{P6112}
P.~Friederich, F.~H{ä}se, J.~Proppe, and A.~Aspuru-Guzik, ``Machine-learned
  potentials for next-generation matter simulations,'' {\em Nat. Mater},
  vol.~20, pp.~750--761, 2021.

\bibitem{P6121}
J.~Behler and G.~Cs\'{a}nyi, ``Machine learning potentials for extended systems
  - a perspective,'' {\em Eur. Phys. J. B}, vol.~94, p.~142, 2021.

\bibitem{P5673}
V.~L. Deringer, M.~A. Caro, and G.~Cs\'{a}nyi, ``Machine learning interatomic
  potentials as emerging tools for materials science,'' {\em Adv. Mater.},
  vol.~31, p.~1902765, 2019.

\bibitem{P5788}
P.~O. Dral, ``Quantum chemistry in the age of machine learning,'' {\em J. Phys.
  Chem. Lett.}, vol.~11, pp.~2336--2347, 2020.

\bibitem{P5793}
F.~No\'e, A.~Tkatchenko, K.-R. M{ü}ller, and C.~Clementi, ``Machine learning
  for molecular simulation,'' {\em Ann. Rev. Phys. Chem.}, vol.~71,
  pp.~361--390, 2020.

\bibitem{P4263}
C.~M. Handley and J.~Behler, ``Next generation interatomic potentials for
  condensed systems,'' {\em Eur. Phys. J. B}, vol.~87, p.~152, 2014.

\bibitem{P6018}
J.~Behler, ``Four generations of high-dimensional neural network potentials,''
  {\em Chem. Rev.}, vol.~121, no.~16, p.~10037–10072, 2021.

\bibitem{P5971}
V.~L. Deringer, A.~P. Bart\'{o}k, N.~Bernstein, D.~M. Wilkins, M.~Ceriotti, and
  G.~Cs\'{a}nyi, ``Gaussian process regression for materials and molecules,''
  {\em Chem. Rev.}, vol.~121, pp.~10073--10141, 2021.

\bibitem{P6131}
E.~Kocer, T.~W. Ko, and J.~Behler, ``Neural network potentials: A concise
  overview of methods,'' {\em Annu. Rev. Phys. Chem.}, vol.~73, pp.~163--186,
  2022.

\bibitem{P5977}
T.~W. Ko, J.~A. Finkler, S.~Goedecker, and J.~Behler, ``General-purpose machine
  learning potentials capturing nonlocal charge transfer,'' {\em Acc. Chem.
  Res.}, vol.~54, pp.~808--817, 2021.

\bibitem{P6631}
S.~K{ä}ser, L.~I. Vazquez-Salazar, M.~Meuwly, and K.~T{ö}pfer, ``Neural
  network potentials for chemistry: concepts, applications and prospects,''
  {\em Digit. Discov.}, vol.~2, p.~28, 2023.

\bibitem{P6548}
A.~M. Tokita and J.~Behler, ``Tutorial: How to train a neural network
  potential,'' {\em J. Chem. Phys.}, vol.~159, p.~121501, 2023.

\bibitem{P0316}
T.~B. Blank, S.~D. Brown, A.~W. Calhoun, and D.~J. Doren, ``Neural network
  models of potential energy surfaces,'' {\em J. Chem. Phys.}, vol.~103,
  pp.~4129--4137, 1995.

\bibitem{P0830}
H.~Gassner, M.~Probst, A.~Lauenstein, and K.~Hermansson, ``Representation of
  intermolecular potential functions by neural networks,'' {\em J. Phys. Chem.
  A}, vol.~102, pp.~4596--4605, 1998.

\bibitem{P0421}
S.~Lorenz, A.~Gro\ss, and M.~Scheffler, ``Representing high-dimensional
  potential-energy surfaces for reactions at surfaces by neural networks,''
  {\em Chem. Phys. Lett.}, vol.~395, pp.~210--215, 2004.

\bibitem{P1388}
J.~Behler, S.~Lorenz, and K.~Reuter, ``Representing molecule-surface
  interactions with symmetry-adapted neural networks,'' {\em J. Chem. Phys.},
  vol.~127, p.~014705, 2007.

\bibitem{P3062}
A.~Brown, B.~J. Braams, K.~Christoffel, Z.~Jin, and J.~M. Bowman, ``Classical
  and quasiclassical spectral analysis of ch$_5^+$ using an ab initio potential
  energy surface,'' {\em J. Chem. Phys.}, vol.~119, pp.~8790--8793, 2003.

\bibitem{P1174}
J.~Behler and M.~Parrinello, ``Generalized neural-network representation of
  high-dimensional potential-energy surfaces,'' {\em Phys. Rev. Lett.},
  vol.~98, p.~146401, 2007.

\bibitem{P5128}
J.~Behler, ``First principles neural network potentials for reactive
  simulations of large molecular and condensed systems,'' {\em Angew. Chem.
  Int. Ed.}, vol.~56, p.~12828, 2017.

\bibitem{P4444}
J.~Behler, ``Constructing high-dimensional neural network potentials: A
  tutorial review,'' {\em Int. J. Quantum Chem.}, vol.~115, pp.~1032--1050,
  2015.

\bibitem{P4106}
J.~Behler, ``Representing potential energy surfaces by high-dimensional neural
  network potentials,'' {\em J. Phys. Condens. Matter}, vol.~26, no.~18,
  p.~183001, 2014.

\bibitem{P2882}
J.~Behler, ``Atom-centered symmetry functions for constructing high-dimensional
  neural network potentials,'' {\em J. Chem. Phys.}, vol.~134, p.~074106, 2011.

\bibitem{BartokPRB2013}
A.~P. Bart\'ok, R.~Kondor, and G.~Cs\'anyi, ``On representing chemical
  environments,'' {\em Phys. Rev. B}, vol.~87, p.~184115, May 2013.

\bibitem{PronobisJCTC2018}
W.~Pronobis, A.~Tkatchenko, and K.-R. Müller, ``Many-body descriptors for
  predicting molecular properties with machine learning: Analysis of pairwise
  and three-body interactions in molecules,'' {\em J. Chem. Theory Comput.},
  vol.~14, no.~6, pp.~2991--3003, 2018.
\newblock PMID: 29750522.

\bibitem{MusilChemRev2021}
F.~Musil, A.~Grisafi, A.~P. Bartók, C.~Ortner, G.~Csányi, and M.~Ceriotti,
  ``Physics-inspired structural representations for molecules and materials,''
  {\em Chem. Rev.}, vol.~121, no.~16, pp.~9759--9815, 2021.

\bibitem{P5794}
R.~Drautz, ``Atomic cluster expansion for accurate and transferable interatomic
  potentials,'' {\em Phys. Rev. B}, vol.~99, p.~014104, 2019.

\bibitem{P4945}
J.~S. Smith, O.~Isayev, and A.~E. Roitberg, ``Ani-1: An extensible neural
  network potential with dft accuracy at force field computational cost,'' {\em
  Chem. Sci.}, vol.~8, pp.~3192--3203, 2017.

\bibitem{P5596}
L.~Zhang, J.~Han, H.~Wang, R.~Car, and W.~E, ``Deep potential molecular
  dynamics: A scalable model with the accuracy of quantum mechanics,'' {\em
  Phys. Rev. Lett.}, vol.~120, p.~143001, 2018.

\bibitem{P5367}
H.~Wang, L.~Zhang, J.~Han, and W.~E, ``Deepmd-kit: A deep learning package for
  many-body potential energy representation and molecular dynamics,'' {\em
  Comput. Phys. Commun.}, vol.~228, pp.~178--184, 2018.

\bibitem{P5076}
J.~Han, L.~Zhang, R.~Car, and W.~E, ``Deep potential: a general representation
  of a many-body potential energy surface,'' {\em Commun. Comput. Phys.},
  vol.~23, pp.~629--639, 2018.

\bibitem{P5792}
Y.~Zhang, C.~Hu, and B.~Jiang, ``Embedded atom neural network potentials:
  Efficient and accurate machine learning with a physically inspired
  representation,'' {\em J. Phys. Chem. Lett.}, vol.~10, pp.~4962--4967, 2019.

\bibitem{P6605}
V.~Zaverkin, D.~Holzm{ü}ller, I.~Steinwart, and J.~K{ä}stner, ``Fast and
  sample-efficient interatomic neural network potentials for molecules and
  materials based on gaussian moments,'' {\em J. Chem. Theory Comput.},
  vol.~17, pp.~6658--6670, 2021.

\bibitem{P5368}
J.~Gilmer, S.~S. Schoenholz, P.~F. Riley, O.~Vinyals, and G.~E. Dahl, ``Neural
  message passing for quantum chemistry,'' in {\em Proceedings of the 34th
  International Conference on Machine Learning} (D.~Precup and Y.~W. Teh,
  eds.), vol.~70 of {\em Proceedings of Machine Learning Research},
  pp.~1263--1272, PMLR, 2017.

\bibitem{P4937}
K.~T. Schütt, F.~Arbabzadah, S.~Chmiela, K.~R. Müller, and A.~Tkatchenko,
  ``Quantum-chemical insights from deep tensor neural networks,'' {\em Nat.
  Commun.}, vol.~8, p.~13890, 2017.

\bibitem{P5366}
K.~T. Schütt, H.~E. Sauceda, P.-J. Kindermans, A.~Tkatchenko, and K.-R.
  Mueller, ``Schnet - a deep learning architecture for molecules and
  materials,'' {\em J. Chem. Phys.}, vol.~148, p.~241722, 2018.

\bibitem{P5817}
R.~Zubatyuk, J.~S. Smith, J.~Leszczynski, and O.~Isayev, ``Accurate and
  transferable multitask prediction of chemical properties with an
  atoms-in-molecules neural network,'' {\em Sci. Adv.}, vol.~5, p.~eaav6490,
  2019.

\bibitem{P6026}
K.~Sch{ü}tt, O.~Unke, and M.~Gastegger, ``Equivariant message passing for the
  prediction of tensorial properties and molecular spectra,'' in {\em
  Proceedings of the 38th International Conference on Machine Learning}
  (M.~Meila and T.~Zhang, eds.), vol.~139 of {\em Proceedings of Machine
  Learning Research}, pp.~9377--9388, PMLR, 18--24 Jul 2021.

\bibitem{P6017}
S.~Batzner, A.~Musaelian, L.~Sun, M.~Geiger, J.~P. Mailoa, M.~Kornbluth,
  N.~Molinari, T.~E. Smidt, and B.~Kozinsky, ``E (3)-equivariant graph neural
  networks for data-efficient and accurate interatomic potentials,'' {\em Nat.
  Commun.}, vol.~13, no.~1, p.~2453, 2022.

\bibitem{P6569}
A.~Musaelian, S.~Batzner, A.~Johansson, L.~Sun, C.~J. Owen, M.~Kornbluth, and
  B.~Kozinsky, ``Learning local equivariant representations for large-scale
  atomistic dynamics,'' {\em Nat. Commun.}, vol.~14, p.~579, 2023.

\bibitem{P6572}
I.~Batatia, D.~P. Kovacs, G.~Simm, C.~Ortner, and G.~Csanyi, ``Mace: Higher
  order equivariant message passing neural networks for fast and accurate force
  fields,'' in {\em Adv. Neural Inf. Process. Syst.} (S.~Koyejo, S.~Mohamed,
  A.~Agarwal, D.~Belgrave, K.~Cho, and A.~Oh, eds.), vol.~35, pp.~11423--11436,
  Curran Associates, Inc., 2022.

\bibitem{P2630}
A.~P. Bart\'{o}k, M.~C. Payne, R.~Kondor, and G.~Cs\'{a}nyi, ``Gaussian
  approximation potentials: The accuracy of quantum mechanics, without the
  electrons,'' {\em Phys. Rev. Lett.}, vol.~104, p.~136403, 2010.

\bibitem{P3885}
A.~P. Bart\'{o}k, R.~Kondor, and G.~Cs\'{a}nyi, ``On representing chemical
  environments,'' {\em Phys. Rev. B}, vol.~87, p.~184115, 2013.

\bibitem{P4862}
A.~V. Shapeev, ``Moment tensor potentials: a class of systematically improvable
  interatomic potentials,'' {\em Multiscale Model. Simul.}, vol.~14,
  pp.~1153--1173, 2016.

\bibitem{P4644}
A.~P. Thompson, L.~P. Swiler, C.~R. Trott, S.~M. Foiles, and G.~J. Tucker,
  ``Spectral neighbor analysis method for automated generation of
  quantum-accurate interatomic potentials,'' {\em J. Comp. Phys.}, vol.~285,
  pp.~316--330, 2015.

\bibitem{P5975}
S.~Yue, M.~C. Muniz, M.~F.~C. Andrade, L.~Zhang, R.~Car, and A.~Z.
  Panagiotopoulos, ``When do short-range atomistic machine-learning models fall
  short?,'' {\em J. Chem. Phys.}, vol.~154, p.~034111, 2021.

\bibitem{P2391}
S.~Houlding, S.~Y. Liem, and P.~L.~A. Popelier, ``A polarizable high-rank
  quantum topological electrostatic potential developed using neural networks:
  Molecular dynamics simulations on the hydrogen fluoride dimer,'' {\em Int. J.
  Quantum Chem.}, vol.~107, pp.~2817--2827, 2007.

\bibitem{P3199}
M.~J.~L. Mills and P.~L.~A. Popelier, ``Polarisable multipolar electrostatics
  from the machine learning method kriging: an application to alanine,'' {\em
  Theor. Chem. Acc.}, vol.~131, p.~1137, 2012.

\bibitem{P2392}
C.~M. Handley, G.~I. Hawe, D.~B. Kell, and P.~L.~A. Popelier, ``Optimal
  construction of a fast and accurate polarisable water potential based on
  multipole moments trained by machine learning,'' {\em Phys. Chem. Chem.
  Phys.}, vol.~11, pp.~6365--6376, 2009.

\bibitem{P2962}
N.~Artrith, T.~Morawietz, and J.~Behler, ``High-dimensional neural-network
  potentials for multicomponent systems: Applications to zinc oxide,'' {\em
  Phys. Rev. B}, vol.~83, p.~153101, 2011.

\bibitem{P3132}
T.~Morawietz, V.~Sharma, and J.~Behler, ``A neural network potential-energy
  surface for the water dimer based on environment-dependent atomic energies
  and charges,'' {\em J. Chem. Phys.}, vol.~136, p.~064103, 2012.

\bibitem{P5313}
K.~Yao, J.~E. Herr, D.~W. Toth, R.~Mckintyre, and J.~Parkhill, ``The
  tensormol-0.1 model chemistry: a neural network augmented with long-range
  physics,'' {\em Chem. Sci.}, vol.~9, pp.~2261--2269, 2018.

\bibitem{P5577}
O.~T. Unke and M.~Meuwly, ``Physnet: A neural network for predicting energies,
  forces, dipole moments, and partial charges,'' {\em J. Chem. Theory Comput.},
  vol.~15, pp.~3678--3693, 2019.

\bibitem{P6056}
O.~T. Unke, S.~Chmiela, M.~Gastegger, K.~T. Schütt, H.~E. Sauceda, and K.-R.
  M{ü}ller, ``Spookynet: Learning force fields with electronic degrees of
  freedom and nonlocal effects,'' {\em Nat. Commun.}, vol.~12, no.~1, p.~7273,
  2021.

\bibitem{P5885}
T.~Bereau, D.~Andrienko, and O.~A. von Lilienfeld, ``Transferable atomic
  multipole machine learning models for small organic molecules,'' {\em J.
  Chem. Theory Comput.}, vol.~11, pp.~3225--3233, 2015.

\bibitem{P5372}
A.~E. Sifain, N.~Lubbers, B.~T. Nebgen, J.~S. Smith, A.~Y. Lokhov, O.~Isayev,
  A.~E. Roitberg, K.~Barros, and S.~Tretiak, ``Discovering a transferable
  charge assignment model using machine learning,'' {\em J. Phys. Chem. Lett.},
  vol.~9, pp.~4495--4501, 2018.

\bibitem{P6200}
L.~Zhang, H.~Wang, M.~C. Muniz, A.~Z. Panagiotopoulos, R.~Car, {\em et~al.},
  ``A deep potential model with long-range electrostatic interactions,'' {\em
  J. Chem. Phys.}, vol.~156, no.~12, 2022.

\bibitem{P5205}
M.~Gastegger, J.~Behler, and P.~Marquetand, ``Machine learning molecular
  dynamics for the simulation of infrared spectra,'' {\em Chem. Sci.}, vol.~8,
  pp.~6924--6935, 2017.

\bibitem{P4419}
S.~A. Ghasemi, A.~Hofstetter, S.~Saha, and S.~Goedecker, ``Interatomic
  potentials for ionic systems with density functional accuracy based on charge
  densities obtained by a neural network,'' {\em Phys. Rev. B}, vol.~92,
  p.~045131, 2015.

\bibitem{P1448}
A.~K. Rappe and W.~A. Goddard, III, ``Charge equilibration for molecular
  dynamics simulations,'' {\em J. Phys. Chem.}, vol.~95, pp.~3358--3363, 1991.

\bibitem{P5859}
X.~Xie, K.~A. Persson, and D.~W. Small, ``Incorporating electronic information
  into machine learning potential energy surfaces via approaching the
  ground-state electronic energy as a function of atom-based electronic
  populations,'' {\em J. Chem. Theory Comput.}, vol.~16, pp.~4256--4270, 2020.

\bibitem{P5932}
T.~W. Ko, J.~A. Finkler, S.~Goedecker, and J.~Behler, ``A fourth-generation
  high-dimensional neural network potential with accurate electrostatics
  including non-local charge transfer,'' {\em Nat. Commun.}, vol.~12, p.~398,
  2021.

\bibitem{P6122}
L.~Jacobson, J.~Stevenson, F.~Ramezanghorbani, D.~Ghoreishi, K.~Leswing,
  E.~Harder, and R.~Abel, ``Transferable neural network potential energy
  surfaces for closed-shell organic molecules: Extension to ions,'' {\em J.
  Chem. Theor. Comp.}, vol.~18, pp.~2354--2366, 2022.

\bibitem{P5842}
C.~Schran, J.~Behler, and D.~Marx, ``Automated fitting of neural network
  potentials at coupled cluster accuracy: Protonated water clusters as testing
  ground,'' {\em J. Chem. Theory Comput.}, vol.~16, pp.~88--99, 2020.

\bibitem{P5399}
J.~S. Smith, B.~Nebgen, N.~Lubbers, O.~Isayev, and A.~E. Roitberg, ``Less is
  more: Sampling chemical space with active learning,'' {\em J. Chem. Phys.},
  vol.~148, p.~241733, 2018.

\bibitem{P5782}
L.~Zhang, D.-Y. Lin, H.~Wang, R.~Car, and W.~E, ``Active learning of uniformly
  accurate interatomic potentials for materials simulation,'' {\em Phys. Rev.
  Mater.}, vol.~3, p.~023804, 2019.

\bibitem{P6058}
C.~Schran, F.~L. Thiemann, P.~Rowe, E.~A. M{ü}ller, O.~Marsalek, and
  A.~Michaelides, ``Machine learning potentials for complex aqueous systems
  made simple,'' {\em Proc. Natl. Acad. Sci. U. S. A.}, vol.~118, no.~38,
  p.~e2110077118, 2021.

\bibitem{P0833}
K.~T. No, B.~H. Chang, S.~Y. Kim, M.~S. Jhon, and H.~A. Scheraga, ``Description
  of the potential energy surface of the water dimer with an artificial neural
  network,'' {\em Chem. Phys. Lett.}, vol.~271, pp.~152--156, 1997.

\bibitem{P0826}
S.~Manzhos, X.~Wang, R.~Dawes, and T.~Carrington, Jr, ``A nested
  molecule-independent neural network approach for high-quality potential
  fits,'' {\em J. Phys. Chem. A}, vol.~110, pp.~5295--5304, 2006.

\bibitem{P3259}
H.~T.~T. Nguyen and H.~M. Le, ``Modified feed-forward neural network structures
  and combined-function-derivative approximations incorporating exchange
  symmetry for potential energy surface fitting,'' {\em J. Phys. Chem. A},
  vol.~116, pp.~4629--4638, 2012.

\bibitem{P3140}
X.~Huang, B.~J. Braams, and J.~M. Bowman, ``Ab initio potential energy and
  dipole moment surfaces for h$_5$o$_2^+$,'' {\em J. Chem. Phys.}, vol.~122,
  p.~044308, 2005.

\bibitem{P3139}
X.~Huang, B.~J. Braams, and J.~M. Bowman, ``Ab initio potential energy and
  dipole moment surfaces of (h$_2$o)$_2$,'' {\em J. Phys. Chem. A}, vol.~110,
  pp.~445--451, 2006.

\bibitem{P6161}
X.~Huang, B.~J. Braams, and J.~M. Bowman, ``New ab initio potential energy
  surface and the vibration-rotation-tunneling levels of (h$_2$o)$_2$ and
  (d$_2$o)$_2$,'' {\em J. Chem. Phys.}, vol.~128, p.~034312, 2008.

\bibitem{P2296}
Y.~Wang, B.~C. Shepler, B.~J. Braams, and J.~M. Bowman, ``Full-dimensional, ab
  initio potential energy and dipole moment surfaces for water,'' {\em J. Chem.
  Phys.}, vol.~131, p.~054511, 2009.

\bibitem{P6160}
A.~Shank, Y.~Wang, A.~Kaledin, B.~J. Braams, and J.~M. Bowman, ``Accurate ab
  initio and “hybrid” potential energy surfaces, intramolecular vibrational
  energies, and classical ir spectrum of the water dimer,'' {\em J. Chem.
  Phys.}, vol.~130, p.~144314, 2009.

\bibitem{P2716}
J.~M. Bowman, B.~J. Braams, S.~Carter, C.~Chen, G.~Czak, B.~Fu, X.~Huang,
  E.~Kamarchik, A.~R. Sharma, B.~C. Shepler, Y.~Wang, and Z.~Xie,
  ``Ab-initio-based potential energy surfaces for complex molecules and
  molecular complexes,'' {\em J. Phys. Chem. Lett.}, vol.~1, pp.~1866--1874,
  2010.

\bibitem{P2720}
Y.~Wang and J.~M. Bowman, ``Towards an ab initio flexible potential for water,
  and post-harmonic quantum vibrational analysis of water clusters,'' {\em
  Chem. Phys. Lett.}, vol.~491, pp.~1--10, 2010.

\bibitem{P3958}
Z.~Xie and J.~M. Bowman, ``Permutationally invariant polynomial basis for
  molecular energy surface fitting via monomial symmetrization,'' {\em J. Chem.
  Theory Comp.}, vol.~6, p.~26, 2010.

\bibitem{P4365}
Y.~Wang, X.~Huang, B.~C. Shepler, B.~J. Braams, and J.~M. Bowman, ``Flexible,
  ab initio potential, and dipole moment surfaces for water. i. tests and
  applications for clusters up to the 22-mer,'' {\em J. Chem. Phys.}, vol.~134,
  p.~094509, 2011.

\bibitem{P5641}
Q.~Yu and J.~M. Bowman, ``Classical, thermostated ring polymer, and quantum
  vscf/vci calculations of ir spectra of h$_7$o$_3^+$ and h$_9$o$_4^+$ (eigen)
  and comparison with experiment,'' {\em J. Phys. Chem. A}, vol.~123,
  pp.~1399--1409, 2019.

\bibitem{P6321}
Q.~Yu, C.~Qu, P.~L. Houston, R.~Conte, A.~Nandi, and J.~M. Bowman, ``q-aqua: A
  many-body ccsd(t) water potential, including fourbody interactions,
  demonstrates the quantum nature of water from clusters to the liquid phase,''
  {\em J. Phys. Chem. Lett.}, vol.~13, pp.~5068--5074, 2022.

\bibitem{P6645}
C.~Qu, Q.~Yu, R.~Conte, P.~L. Houston, A.~Nandi, and J.~M. Bomwan, ``A
  $\delta$-machine learning approach for force fields, illustrated by a ccsd
  (t) 4-body correction to the mb-pol water potential,'' {\em Digit. Discov.},
  vol.~1, no.~5, pp.~658--664, 2022.

\bibitem{P3875}
T.~Morawietz and J.~Behler, ``A density-functional theory-based neural network
  potential for water clusters including van der waals corrections,'' {\em J.
  Phys. Chem. A}, vol.~117, p.~7356, 2013.

\bibitem{P3971}
T.~Morawietz and J.~Behler, ``A full-dimensional neural network
  potential-energy surface for water clusters up to the hexamer,'' {\em Z.
  Phys. Chem.}, vol.~227, pp.~1559--1581, 2013.

\bibitem{P4472}
S.~K. Natarajan, T.~Morawietz, and J.~Behler, ``Representing the
  potential-energy surface of protonated water clusters by high-dimensional
  neural network potentials,'' {\em Phys. Chem. Chem. Phys.}, vol.~17, p.~8356,
  2015.

\bibitem{P6646}
H.~Zhou, Y.-J. Feng, C.~Wang, T.~Huang, Y.-R. Liu, S.~Jiang, C.-Y. Wang, and
  W.~Huang, ``A high-accuracy machine-learning water model for exploring water
  nanocluster structures,'' {\em Nanoscale}, vol.~13, no.~28, pp.~12212--12222,
  2021.

\bibitem{P5504}
C.~Schran, F.~Brieuc, and D.~Marx, ``Converged colored noise path integral
  molecular dynamics study of the zundel cation down to ultralow temperatures
  at coupled cluster accuracy,'' {\em J. Chem. Theory Comput.}, vol.~14,
  pp.~5068--5078, 2018.

\bibitem{P5882}
C.~Schran and D.~Marx, ``Quantum nature of the hydrogen bond from ambient
  conditions down to ultra-low temperatures,'' {\em Phys. Chem. Chem. Phys.},
  vol.~21, pp.~24967--24975, 2019.

\bibitem{P6136}
C.~Schran, F.~Brieuc, and D.~Marx, ``Transferability of machine learning
  potentials: Protonated water neural network potential applied to the
  protonated water hexamer,'' {\em J. Chem. Phys.}, vol.~154, p.~051101, 2021.

\bibitem{P6223}
R.~Beckmann, F.~Brieuc, C.~Schran, and D.~Marx, ``Infrared spectra at coupled
  cluster accuracy from neural network representations,'' {\em J. Chem. Theory
  Comput.}, vol.~18, no.~9, pp.~5492--5501, 2022.

\bibitem{P5305}
C.~Schran, F.~Uhl, J.~Behler, and D.~Marx, ``High-dimensional neural network
  potentials for solvation: The case of protonated water clusters in helium,''
  {\em J. Chem. Phys.}, vol.~148, p.~102310, 2017.

\bibitem{P2211}
C.~M. Handley and P.~L.~A. Popelier, ``Dynamically polarizable water potential
  based on multipole moments trained by machine learning,'' {\em J. Chem.
  Theory Comput.}, vol.~5, pp.~1474--1489, 2009.

\bibitem{P6632}
S.~J. Davie, N.~Di~Pasquale, and P.~L. Popelier, ``Incorporation of local
  structure into kriging models for the prediction of atomistic properties in
  the water decamer,'' {\em J. Comput. Chem.}, vol.~37, no.~27, pp.~2409--2422,
  2016.

\bibitem{P6634}
Z.~E. Hughes, E.~Ren, J.~C. Thacker, B.~C. Symons, A.~F. Silva, and P.~L.
  Popelier, ``A fflux water model: Flexible, polarizable and with a multipolar
  description of electrostatics,'' {\em J. Comput. Chem.}, vol.~41, no.~7,
  pp.~619--628, 2020.

\bibitem{P5881}
A.~Grisafi, D.~M. Wilkins, G.~Cs{\'a}nyi, and M.~Ceriotti, ``Symmetry-adapted
  machine learning for tensorial properties of atomistic systems,'' {\em Phys.
  Rev. Lett.}, vol.~120, no.~3, p.~036002, 2018.

\bibitem{P6633}
V.~H.~A. Nguyen and A.~Lunghi, ``Predicting tensorial molecular properties with
  equivariant machine learning models,'' {\em Phys. Rev. B}, vol.~105, no.~16,
  p.~165131, 2022.

\bibitem{P6635}
J.~Schuhmacher, G.~Mazzola, F.~Tacchino, O.~Dmitriyeva, T.~Bui, S.~Huang, and
  I.~Tavernelli, ``Extending the reach of quantum computing for materials
  science with machine learning potentials,'' {\em AIP Advances}, vol.~12,
  no.~11, 2022.

\bibitem{P6636}
S.~Bose, D.~Dhawan, S.~Nandi, R.~R. Sarkar, and D.~Ghosh, ``Machine learning
  prediction of interaction energies in rigid water clusters,'' {\em Phys.
  Chem. Chem. Phys.}, vol.~20, no.~35, pp.~22987--22996, 2018.

\bibitem{P6637}
A.~F. Silva, L.~J. Duarte, and P.~L. Popelier, ``Contributions of iqa electron
  correlation in understanding the chemical bond and non-covalent
  interactions,'' {\em Struct. Chem.}, vol.~31, no.~2, pp.~507--519, 2020.

\bibitem{P6638}
L.~Cheng, J.~Sun, J.~E. Deustua, V.~C. Bhethanabotla, and T.~F. Miller,
  ``Molecular-orbital-based machine learning for open-shell and multi-reference
  systems with kernel addition gaussian process regression,'' {\em J. Chem.
  Phys.}, vol.~157, no.~15, 2022.

\bibitem{P5789}
L.~Cheng, M.~Welborn, A.~S. Christensen, and T.~F. Miller, ``A universal
  density matrix functional from molecular orbital-based machine learning:
  Transferability across organic molecules,'' {\em J. Chem. Phys.}, vol.~150,
  no.~13, p.~131103, 2019.

\bibitem{P6639}
J.~P. Coe, ``Machine learning configuration interaction,'' {\em J. Chem. Theory
  Comput.}, vol.~14, no.~11, pp.~5739--5749, 2018.

\bibitem{P6640}
J.~P. Coe, ``Machine learning configuration interaction for ab initio potential
  energy curves,'' {\em J. Chem. Theory Comput.}, vol.~15, no.~11,
  pp.~6179--6189, 2019.

\bibitem{P6641}
F.~Lu, L.~Cheng, R.~J. DiRisio, J.~M. Finney, M.~A. Boyer, P.~Moonkaen, J.~Sun,
  S.~J. Lee, J.~E. Deustua, T.~F. Miller~III, {\em et~al.}, ``Fast near ab
  initio potential energy surfaces using machine learning,'' {\em J. Phys.
  Chem. A}, vol.~126, no.~25, pp.~4013--4024, 2022.

\bibitem{P6077}
M.~Welborn, L.~Cheng, and T.~F. Miller~III, ``Transferability in machine
  learning for electronic structure via the molecular orbital basis,'' {\em J.
  Chem. Theory Comput.}, vol.~14, no.~9, pp.~4772--4779, 2018.

\bibitem{P6642}
R.~J. DiRisio, F.~Lu, and A.~B. McCoy, ``Gpu-accelerated neural network
  potential energy surfaces for diffusion monte carlo,'' {\em J. Phys. Chem.
  A}, vol.~125, no.~26, pp.~5849--5859, 2021.

\bibitem{P0838}
K.-W. Cho, K.~T. No, and H.~A. Scheraga, ``A polarizable force field for water
  using an artificial neural network,'' {\em J. Mol. Struct.}, vol.~641,
  pp.~77--91, 2002.

\bibitem{P1472}
W.~L. Jorgensen, J.~Chandrasekhar, J.~D. Madura, R.~W. Impey, and M.~L. Klein,
  ``Comparison of simple potential functions for simulating liquid water,''
  {\em J. Chem. Phys.}, vol.~79, pp.~926--935, 1983.

\bibitem{P3872}
A.~P. Bart\'{o}k, M.~J. Gillan, F.~R. Manby, and G.~Cs\'{a}nyi,
  ``Machine-learning approach for one- and two-body corrections to density
  functional theory: Applications to molecular and condensed water,'' {\em
  Phys. Rev. B}, vol.~88, p.~054104, 2013.

\bibitem{P3886}
D.~Alf\`{e}, A.~P. Bart\'{o}k, G.~Cs\'{a}nyi, and M.~J. Gillan, ``Energy
  benchmarking with quantum monte carlo for water nano-droplets and bulk liquid
  water,'' {\em J. Chem. Phys.}, vol.~128, p.~221102, 2013.

\bibitem{P4556}
T.~Morawietz, A.~Singraber, C.~Dellago, and J.~Behler, ``How van der waals
  interactions determine the unique properties of water,'' {\em Proc. Natl.
  Acad. Sci. U. S. A.}, vol.~113, p.~8368, 2016.

\bibitem{P5711}
A.~Singraber, R.~Morawietz, J.~Behler, and C.~Dellago, ``Density anomaly of
  water at negative pressures from first principles,'' {\em J. Phys.: Condens.
  Matter}, vol.~30, p.~254005, 2018.

\bibitem{montero2023kinetics}
P.~Montero~de Hijes, S.~Romano, A.~Gorfer, and C.~Dellago, ``The kinetics of
  the ice--water interface from ab initio machine learning simulations,'' {\em
  J. Chem. Phys.}, vol.~158, no.~20, 2023.

\bibitem{P4586}
B.~Cheng, J.~Behler, and M.~Ceriotti, ``Nuclear quantum effects in water at the
  triple point: Using theory as a link between experiments,'' {\em J. Phys.
  Chem. Lett.}, vol.~7, pp.~2210--2215, 2016.

\bibitem{P4971}
V.~Kapil, J.~Behler, and M.~Ceriotti, ``High order path integrals made easy,''
  {\em J. Chem. Phys.}, vol.~145, p.~234103, 2016.

\bibitem{P6298}
V.~Kapil, D.~M. Wilkins, J.~Lan, and M.~Ceriotti, ``Inexpensive modeling of
  quantum dynamics using path integral generalized langevin equation
  thermostats,'' {\em J. Chem. Phys.}, vol.~152, p.~124104, 2020.

\bibitem{yao2020temperature}
Y.~Yao and Y.~Kanai, ``Temperature dependence of nuclear quantum effects on
  liquid water via artificial neural network model based on scan meta-gga
  functional,'' {\em J. Chem. Phys.}, vol.~153, no.~4, 2020.

\bibitem{P6217}
C.~Li and G.~A. Voth, ``Using machine learning to greatly accelerate path
  integral ab initio molecular dynamics,'' {\em J. Chem. Theor. Comp.},
  vol.~18, pp.~599--604, 2022.

\bibitem{P5479}
B.~Cheng, E.~A. Engel, J.~Behler, C.~Dellago, and M.~Ceriotti, ``Ab initio
  thermodynamics of liquid and solid water,'' {\em Proc. Natl. Acad. Sci. U. S.
  A.}, vol.~116, pp.~1110--1115, 2019.

\bibitem{yao2021nuclear}
Y.~Yao and Y.~Kanai, ``Nuclear quantum effect and its temperature dependence in
  liquid water from random phase approximation via artificial neural network,''
  {\em J. Phys. Chem. Lett.}, vol.~12, no.~27, pp.~6354--6362, 2021.

\bibitem{P5570}
T.~Morawietz, O.~Marsalek, S.~R. Pattenaude, L.~M. Streacker, D.~Ben-Amotz, and
  T.~E. Markland, ``The interplay of structure and dynamics in the raman
  spectrum of liquid water over the full frequency and temperature range,''
  {\em J. Phys. Chem. Lett.}, vol.~9, pp.~851--857, 2018.

\bibitem{P5639}
T.~Morawietz, A.~S. Urbina, P.~K. Wise, X.~Wu, W.~Lu, D.~Ben-Amotz, and T.~E.
  Markland, ``Hiding in the crowd: Spectral signatures of overcoordinated
  hydrogen bond environments,'' {\em J. Phys. Chem. Lett.}, vol.~10,
  pp.~6067--6073, 2019.

\bibitem{zhang2020efficient}
Y.~Zhang, S.~Ye, J.~Zhang, C.~Hu, J.~Jiang, and B.~Jiang, ``Efficient and
  accurate simulations of vibrational and electronic spectra with
  symmetry-preserving neural network models for tensorial properties,'' {\em J.
  Phys. Chem. B}, vol.~124, no.~33, pp.~7284--7290, 2020.

\bibitem{P6283}
L.~Zhang, H.~Wang, and W.~E, ``Adaptive coupling of a deep neural network
  potential to a classical force field,'' {\em J. Chem. Phys.}, vol.~149,
  p.~154107, 2018.

\bibitem{P6280}
L.~Zhang, J.~Han, H.~Wang, R.~Car, and W.~E, ``Deepcg: Constructing
  coarse-grained models via deep neural networks,'' {\em J. Chem. Phys.},
  vol.~149, p.~034101, 2018.

\bibitem{P6282}
C.~Andreani, G.~Romanelli, A.~Parmentier, R.~Senesi, A.~I. Kolesnikov, H.-Y.
  Ko, M.~F.~C. Andrade, and R.~Car, ``Hydrogen dynamics in supercritical water
  probed by neutron scattering and computer simulations,'' {\em J. Phys. Chem.
  Lett.}, vol.~11, pp.~9461--9467, 2020.

\bibitem{P6268}
H.-Y. Ko, L.~Zhang, B.~Santra, H.~Wang, W.~E, R.~A.~D. Jr, and R.~Car,
  ``Isotope effects in liquid water via deep potential molecular dynamics,''
  {\em Mol. Phys.}, vol.~117, no.~22, pp.~3269--3281, 2019.

\bibitem{P6291}
J.~Xu, C.~Zhang, L.~Zhang, M.~Chen, B.~Santra, and X.~Wu, ``Isotope effects in
  molecular structures and electronic properties of liquid water via deep
  potential molecular dynamics based on the scan functional,'' {\em Phys. Rev.
  B}, vol.~102, p.~214113, 2020.

\bibitem{liu2022toward}
J.~Liu, J.~Lan, and X.~He, ``Toward high-level machine learning potential for
  water based on quantum fragmentation and neural networks,'' {\em J. Phys.
  Chem. A}, vol.~126, no.~24, pp.~3926--3936, 2022.

\bibitem{P6196}
G.~M. Sommers, M.~F. Calegari~Andrade, L.~Zhang, H.~Wang, and R.~Car, ``Raman
  spectrum and polarizability of liquid water from deep neural networks,'' {\em
  Phys. Chem. Chem. Phys.}, vol.~22, pp.~10592--10602, 2020.

\bibitem{P6260}
L.~Zhang, M.~Chen, X.~Wu, H.~Wang, W.~E, and R.~Car, ``Deep neural network for
  the dielectric response of insulators,'' {\em Phys. Rev. B}, vol.~102,
  p.~041121, Jul 2020.

\bibitem{P6261}
C.~Zhang, F.~Tang, M.~Chen, J.~Xu, L.~Zhang, D.~Y. Qiu, J.~P. Perdew, M.~L.
  Klein, and X.~Wu, ``Modeling liquid water by climbing up jacob’s ladder in
  density functional theory facilitated by using deep neural network
  potentials,'' {\em J. Phys. Chem. B}, vol.~125, pp.~11444--11456, 2021.

\bibitem{P6063}
A.~Krishnamoorthy, K.-i. Nomura, N.~Baradwaj, K.~Shimamura, P.~Rajak,
  A.~Mishra, S.~Fukushima, F.~Shimojo, R.~Kalia, A.~Nakano, and P.~Vashishta,
  ``Dielectric constant of liquid water determined with neural network quantum
  molecular dynamics,'' {\em Phys. Rev. Lett.}, vol.~126, p.~216403, May 2021.

\bibitem{P6286}
Y.~Shi, C.~C. Doyle, and T.~L. Beck, ``Condensed phase water molecular
  multipole moments from deep neural network models trained on ab initio
  simulation data,'' {\em J. Phys. Chem. Lett.}, vol.~12, pp.~10310--10317,
  2021.

\bibitem{schienbein2023spectroscopy}
P.~Schienbein, ``Spectroscopy from machine learning by accurately representing
  the atomic polar tensor,'' {\em J. Chem. Theory Comput.}, vol.~19, no.~3,
  pp.~705--712, 2023.

\bibitem{P6521}
K.~Inoue, Y.~Litman, D.~M. Wilkins, Y.~Nagata, and M.~Okuno, ``Is unified
  understanding of vibrational coupling of water possible? hyper-raman
  measurement and machine learning spectra,'' {\em J. Phys. Chem. Lett.},
  vol.~14, no.~12, pp.~3063--3068, 2023.

\bibitem{P6278}
P.~M. Piaggi, A.~Z. Panagiotopoulos, P.~G. Debenedetti, and R.~Car, ``Phase
  equilibrium of water with hexagonal and cubic ice using the scan
  functional,'' {\em J. Chem. Theory Comp.}, vol.~17, pp.~3065--3077, 2021.

\bibitem{P6151}
L.~Zhang, H.~Wang, R.~Car, and W.~E, ``Phase diagram of a deep potential water
  model,'' {\em Phys. Rev. Lett.}, vol.~126, p.~236001, Jun 2021.

\bibitem{P6195}
A.~Reinhardt and B.~Cheng, ``Quantum-mechanical exploration of the phase
  diagram of water,'' {\em Nat. Commun.}, vol.~12, no.~1, p.~588, 2021.

\bibitem{P6264}
B.~Cheng, M.~Bethkenhagen, C.~J. Pickard, and S.~Hamel, ``Phase behaviours of
  superionic water at planetary conditions,'' {\em Nature Physics}, vol.~17,
  no.~11, pp.~1228--1232, 2021.

\bibitem{P6281}
D.~Tisi, L.~Zhang, R.~Bertossa, H.~Wang, R.~Car, and S.~Baroni, ``Heat
  transport in liquid water from first-principles and deep neural network
  simulations,'' {\em Phys. Rev. B}, vol.~104, p.~224202, 2021.

\bibitem{xu2023accurate}
K.~Xu, Y.~Hao, T.~Liang, P.~Ying, J.~Xu, J.~Wu, and Z.~Fan, ``Accurate
  prediction of heat conductivity of water by a neuroevolution potential,''
  {\em J. Chem. Phys.}, vol.~158, no.~20, 2023.

\bibitem{P6285}
C.~Malosso, L.~Zhang, R.~Car, S.~Baroni, and D.~Tisi, ``Viscosity in water from
  first-principles and deep-neural-network simulations,'' {\em Npj Comput.
  Mater.}, vol.~8, no.~1, p.~139, 2022.

\bibitem{chen2023thermodynamics}
Z.~Chen, M.~L. Berrens, K.-T. Chan, Z.~Fan, and D.~Donadio, ``Thermodynamics of
  water and ice from a fast and scalable first-principles neuroevolution
  potential,'' {\em J. Chem. Eng. Data}, 2023.

\bibitem{P5960}
D.~Lu, H.~Wang, M.~Chen, J.~Liu, L.~Lin, R.~Car, W.~E, W.~Jia, and L.~Zhang,
  ``86 pflops deep potential molecular dynamics simulation of 100 million atoms
  with ab initio accuracy,'' {\em Comp. Phys. Comm.}, vol.~259, p.~107624,
  2021.

\bibitem{bai2005test}
X.-M. Bai and M.~Li, ``Test of classical nucleation theory via
  molecular-dynamics simulation,'' {\em J. Chem. Phys.}, vol.~122, no.~22,
  2005.

\bibitem{sanz2013homogeneous}
E.~Sanz, C.~Vega, J.~Espinosa, R.~Caballero-Bernal, J.~Abascal, and
  C.~Valeriani, ``Homogeneous ice nucleation at moderate supercooling from
  molecular simulation,'' {\em J. Am. Chem. Soc.}, vol.~135, no.~40,
  pp.~15008--15017, 2013.

\bibitem{P6284}
P.~M. Piaggi, J.~Weis, A.~Z. Panagiotopoulos, P.~G. Debenedetti, and R.~Car,
  ``Homogeneous ice nucleation in an ab initio machine-learning model of
  water,'' {\em Proc. Natl. Acad. Sci. U. S. A.}, vol.~119, no.~33,
  p.~e2207294119, 2022.

\bibitem{chen2023imperfectly}
M.~Chen, L.~Tan, H.~Wang, L.~Zhang, and H.~Niu, ``Imperfectly coordinated water
  molecules pave the way for homogeneous ice nucleation,'' {\em arXiv preprint
  arXiv:2304.12665}, 2023.

\bibitem{P6273}
T.~E. Gartner, L.~Zhang, P.~M. Piaggi, R.~Car, A.~Z. Panagiotopoulos, and P.~G.
  Debenedetti, ``Signatures of a liquid{\textendash}liquid transition in an ab
  initio deep neural network model for water,'' {\em Proc. Natl. Acad. Sci. U.
  S. A.}, vol.~117, no.~42, pp.~26040--26046, 2020.

\bibitem{gartner2022liquid}
T.~E. Gartner~III, P.~M. Piaggi, R.~Car, A.~Z. Panagiotopoulos, and P.~G.
  Debenedetti, ``Liquid-liquid transition in water from first principles,''
  {\em Phys. Rev. Lett.}, vol.~129, no.~25, p.~255702, 2022.

\bibitem{piaggi2023meltings}
P.~M. Piaggi, I.~Gartner, Thomas~E., R.~Car, and P.~G. Debenedetti, ``{Melting
  curves of ice polymorphs in the vicinity of the liquid–liquid critical
  point},'' {\em J. Chem. Phys.}, vol.~159, p.~054502, 08 2023.

\bibitem{P6266}
P.~M. Piaggi and R.~Car, ``Enhancing the formation of ionic defects to study
  the ice ih/xi transition with molecular dynamics simulations,'' {\em Mol.
  Phys.}, vol.~0, no.~0, p.~e1916634, 2021.

\bibitem{atsango2023developing}
A.~O. Atsango, T.~Morawietz, O.~Marsalek, and T.~E. Markland, ``Developing
  machine-learned potentials to simultaneously capture the dynamics of excess
  protons and hydroxide ions in classical and path integral simulations,'' {\em
  J. Chem. Phys.}, vol.~159, no.~7, 2023.

\bibitem{liu2023mechanistic}
L.~Liu, Y.~Tian, X.~Yang, and C.~Liu, ``Mechanistic insights into water
  autoionization through metadynamics simulation enhanced by machine
  learning,'' {\em Phys. Rev. Lett.}, vol.~131, no.~15, p.~158001, 2023.

\bibitem{calegari2023probing}
M.~Calegari~Andrade, R.~Car, and A.~Selloni, ``Probing the self-ionization of
  liquid water with ab initio deep potential molecular dynamics,'' {\em Proc.
  Natl. Acad. Sci. U. S. A.}, vol.~120, no.~46, p.~e2302468120, 2023.

\bibitem{P3964}
P.~Geiger and C.~Dellago, ``Neural networks for local structure detection in
  polymorphic systems,'' {\em J. Chem. Phys.}, vol.~139, p.~164105, 2013.

\bibitem{P5832}
M.~Fulford, M.~Salvalaglio, and C.~Molteni, ``Deepice: A deep neural network
  approach to identify ice and water molecules,'' {\em J. Chem. Inf. Model.},
  vol.~59, pp.~2141--2149, 2019.

\bibitem{huang2022machine}
J.~Huang, G.~Huang, and S.~Li, ``A machine learning model to classify dynamic
  processes in liquid water,'' {\em ChemPhysChem}, vol.~23, no.~1,
  p.~e202100599, 2022.

\bibitem{P5968}
B.~Monserrat, J.~G. Brandenburg, E.~A. Engel, and B.~Cheng, ``Liquid water
  contains the building blocks of diverse ice phases,'' {\em Nat. Commun.},
  vol.~11, no.~1, p.~5757, 2020.

\bibitem{guidarelli2023neural}
F.~Guidarelli~Mattioli, F.~Sciortino, and J.~Russo, ``Are neural network
  potentials trained on liquid states transferable to crystal nucleation? a
  test on ice nucleation in the mw water model,'' {\em J. Phys. Chem. B},
  vol.~127, no.~17, pp.~3894--3901, 2023.

\bibitem{P6265}
A.~Torres, L.~S. Pedroza, M.~Fernandez-Serra, and A.~R. Rocha, ``Using neural
  network force fields to ascertain the quality of ab initio simulations of
  liquid water,'' {\em J. Phys. Chem. B}, vol.~125, pp.~10772--10778, 2021.

\bibitem{montero2023MLPs}
P.~Montero~de Hijes, C.~Dellago, R.~Jinnouchi, B.~Schmiedmayer, and G.~Kresse,
  ``Comparing machine learning potentials for water: Kernel-based regression
  and behler-parrinello neural networks,'' {\em arXiv preprint
  arXiv:2312.15213}, 2023.

\bibitem{gomes2023size}
M.~S. Gomes-Filho, A.~Torres, A.~Reily~Rocha, and L.~S. Pedroza, ``Size and
  quality of quantum mechanical data set for training neural network force
  fields for liquid water,'' {\em J. Phys. Chem. B}, vol.~127, no.~6,
  pp.~1422--1428, 2023.

\bibitem{li2022graph}
Z.~Li, K.~Meidani, P.~Yadav, and A.~Barati~Farimani, ``Graph neural networks
  accelerated molecular dynamics,'' {\em J. Chem. Phys.}, vol.~156, no.~14,
  2022.

\bibitem{P5398}
G.~Imbalzano, A.~Anelli, D.~Giofre, S.~Klees, J.~Behler, and M.~Ceriotti,
  ``Automatic selection of atomic fingerprints and reference configurations for
  machine-learning potentials,'' {\em J. Chem. Phys.}, vol.~148, p.~241730,
  2018.

\bibitem{guidarelli2023finger}
F.~Guidarelli~Mattioli, F.~Sciortino, and J.~Russo, ``A neural network
  potential with self-trained atomic fingerprints: a test with the mw water
  potential,'' {\em J. Chem. Phys.}, vol.~158, no.~10, 2023.

\bibitem{young2021transferable}
T.~A. Young, T.~Johnston-Wood, V.~L. Deringer, and F.~Duarte, ``A transferable
  active-learning strategy for reactive molecular force fields,'' {\em Chem.
  Sci.}, vol.~12, no.~32, pp.~10944--10955, 2021.

\bibitem{P5633}
T.~K. Patra, T.~D. Loeffler, H.~Chan, M.~J. Cherukara, B.~Narayanan, and S.~K.
  R.~S. Sankaranarayanan, ``A coarse-grained deep neural network model for
  liquid water,'' {\em Appl. Phys. Lett.}, vol.~115, p.~193101, 2019.

\bibitem{loeffler2020active}
T.~D. Loeffler, T.~K. Patra, H.~Chan, and S.~K. Sankaranarayanan, ``Active
  learning a coarse-grained neural network model for bulk water from sparse
  training data,'' {\em Mol. Syst. Des. Eng.}, vol.~5, no.~5, pp.~902--910,
  2020.

\bibitem{scherer2020kernel}
C.~Scherer, R.~Scheid, D.~Andrienko, and T.~Bereau, ``Kernel-based machine
  learning for efficient simulations of molecular liquids,'' {\em J. Chem.
  Theory Comput.}, vol.~16, no.~5, pp.~3194--3204, 2020.

\bibitem{thaler2021learning}
S.~Thaler and J.~Zavadlav, ``Learning neural network potentials from
  experimental data via differentiable trajectory reweighting,'' {\em Nat.
  Commun.}, vol.~12, no.~1, p.~6884, 2021.

\bibitem{musil2022quantum}
F.~Musil, I.~Zaporozhets, F.~No{\'e}, C.~Clementi, and V.~Kapil, ``Quantum
  dynamics using path integral coarse-graining,'' {\em J. Chem. Phys.},
  vol.~157, no.~18, 2022.

\bibitem{loose2023coarse}
T.~D. Loose, P.~G. Sahrmann, T.~S. Qu, and G.~A. Voth, ``Coarse-graining with
  equivariant neural networks: A path toward accurate and data-efficient
  models,'' {\em J. Phys. Chem. B}, 2023.

\bibitem{P6194}
H.~Chan, M.~J. Cherukara, B.~Narayanan, T.~D. Loeffler, C.~Benmore, S.~K. Gray,
  and S.~K. R.~S. Sankaranarayanan, ``Machine learning coarse grained models
  for water,'' {\em Nat. Commun.}, vol.~10, no.~1, p.~379, 2019.

\bibitem{P6267}
T.~D. Loeffler, H.~Chan, K.~Sasikumar, B.~Narayanan, M.~J. Cherukara, S.~Gray,
  and S.~K. R.~S. Sankaranarayanan, ``Teaching an old dog new tricks: Machine
  learning an improved tip3p potential model for liquid–vapor phase
  phenomena,'' {\em J. Phys. Chem. C}, vol.~123, no.~36, pp.~22643--22655,
  2019.

\bibitem{ye2021machine}
H.-f. Ye, J.~Wang, Y.-g. Zheng, H.-w. Zhang, and Z.~Chen, ``Machine learning
  for reparameterization of four-site water models: Tip4p-bg and tip4p-bgt,''
  {\em Phys. Chem. Chem. Phys.}, vol.~23, no.~17, pp.~10164--10173, 2021.

\bibitem{wang2022machine}
J.~Wang, Y.~Zheng, H.~Zhang, and H.~Ye, ``Machine learning-generated tip4p-bgwt
  model for liquid and supercooled water,'' {\em J. Mol. Liq.}, vol.~367,
  p.~120459, 2022.

\bibitem{han2023incorporating}
B.~Han, C.~M. Isborn, and L.~Shi, ``Incorporating polarization and charge
  transfer into a point-charge model for water using machine learning,'' {\em
  J. Phys. Chem. Lett.}, vol.~14, no.~16, pp.~3869--3877, 2023.

\bibitem{P6492}
Y.~Zhai, A.~Caruso, S.~L. Bore, Z.~Luo, and F.~Paesani, ``A “short blanket”
  dilemma for a state-of-the-art neural network potential for water:
  Reproducing experimental properties or the physics of the underlying
  many-body interactions?,'' {\em J. Chem. Phys.}, vol.~158, p.~084111, 2023.

\bibitem{muniz2023neural}
M.~C. Muniz, R.~Car, and A.~Z. Panagiotopoulos, ``Neural network water model
  based on the mb-pol many-body potential,'' {\em J. Phys. Chem. B}, 2023.

\bibitem{thaler2023scalable}
S.~Thaler, G.~Doehner, and J.~Zavadlav, ``Scalable bayesian uncertainty
  quantification for neural network potentials: Promise and pitfalls,'' {\em J.
  Chem. Theory Comput.}, 2023.

\bibitem{P5360}
T.~T. Nguyen, E.~Székely, G.~Imbalzano, J.~Behler, G.~Csányi, M.~Ceriotti,
  A.~W. Goetz, and F.~Paesani, ``Comparison of permutationally invariant
  polynomials, neural networks, and gaussian approximation potentials in
  representing water interactions through many-body expansions,'' {\em J. Chem.
  Phys.}, vol.~124, p.~241725, 2018.

\bibitem{P5870}
H.~Wang and W.~Yang, ``Force field for water based on neural network,'' {\em J.
  Phys. Chem. Lett.}, vol.~9, pp.~3232--3240, 2018.

\bibitem{P6647}
L.~Yang, J.~Li, F.~Chen, and K.~Yu, ``A transferrable range-separated force
  field for water: Combining the power of both physically-motivated models and
  machine learning techniques,'' {\em J. Chem. Phys.}, vol.~157, no.~21, 2022.

\bibitem{P6347}
B.~C. Symons and P.~L. Popelier, ``Application of quantum chemical topology
  force field fflux to condensed matter simulations: liquid water,'' {\em J.
  Chem. Theory Comput.}, vol.~18, no.~9, pp.~5577--5588, 2022.

\bibitem{P6648}
A.~Konovalov, B.~C. Symons, and P.~L. Popelier, ``On the many-body nature of
  intramolecular forces in fflux and its implications,'' {\em J. Comput.
  Chem.}, vol.~42, no.~2, pp.~107--116, 2021.

\bibitem{P6274}
V.~Zaverkin, D.~Holzmüller, R.~Schuldt, and J.~Kästner, ``Predicting
  properties of periodic systems from cluster data: A case study of liquid
  water,'' {\em J. Chem. Phys.}, vol.~156, p.~114103, 2022.

\bibitem{P6360}
J.~Daru, H.~Forbert, J.~Behler, and D.~Marx, ``Coupled cluster molecular
  dynamics of condensed phase systems enabled by machine learning potentials:
  Liquid water benchmark,'' {\em Phys. Rev. Lett.}, vol.~129, p.~226001, 2022.

\bibitem{P6362}
M.~S. Chen, J.~Lee, H.-Z. Ye, T.~C. Berkelbach, D.~R. Reichman, and T.~E.
  Markland, ``Data-efficient machine learning potentials from transfer learning
  of periodic correlated electronic structure methods: Liquid water at afqmc,
  ccsd, and ccsd (t) accuracy,'' {\em J. Chem. Theory Comput.}, 2023.

\bibitem{grisafi2019incorporating}
A.~Grisafi and M.~Ceriotti, ``Incorporating long-range physics in atomic-scale
  machine learning,'' {\em J. Chem. Phys.}, vol.~151, no.~20, 2019.

\bibitem{P6272}
A.~Gao and R.~C. Remsing, ``Self-consistent determination of long-range
  electrostatics in neural network potentials,'' {\em Nat. Commun.}, vol.~13,
  no.~1, p.~1572, 2022.

\bibitem{dhattarwal2023dielectric}
H.~S. Dhattarwal, A.~Gao, and R.~C. Remsing, ``Dielectric saturation in water
  from a long-range machine learning model,'' {\em J. Phys. Chem. B}, vol.~127,
  no.~16, pp.~3663--3671, 2023.

\bibitem{ple2023force}
T.~Pl{\'e}, L.~Lagard{\`e}re, and J.-P. Piquemal, ``Force-field-enhanced neural
  network interactions: from local equivariant embedding to atom-in-molecule
  properties and long-range effects,'' {\em arXiv preprint arXiv:2301.08734},
  2023.

\bibitem{kovacs2023evaluation}
D.~P. Kovacs, I.~Batatia, E.~S. Arany, and G.~Csanyi, ``Evaluation of the mace
  force field architecture: from medicinal chemistry to materials science,''
  {\em arXiv preprint arXiv:2305.14247}, 2023.

\bibitem{kovacs2023mace}
D.~P. Kov{\'a}cs, J.~H. Moore, N.~J. Browning, I.~Batatia, J.~T. Horton,
  V.~Kapil, I.-B. Magd{\u{a}}u, D.~J. Cole, and G.~Cs{\'a}nyi, ``Mace-off23:
  Transferable machine learning force fields for organic molecules,'' {\em
  arXiv preprint arXiv:2312.15211}, 2023.

\bibitem{batatia2023foundation}
I.~Batatia, P.~Benner, Y.~Chiang, A.~M. Elena, D.~P. Kov{\'a}cs, J.~Riebesell,
  X.~R. Advincula, M.~Asta, W.~J. Baldwin, N.~Bernstein, {\em et~al.}, ``A
  foundation model for atomistic materials chemistry,'' {\em arXiv preprint
  arXiv:2401.00096}, 2023.

\bibitem{fu2022forces}
X.~Fu, Z.~Wu, W.~Wang, T.~Xie, S.~Keten, R.~Gomez-Bombarelli, and T.~Jaakkola,
  ``Forces are not enough: Benchmark and critical evaluation for machine
  learning force fields with molecular simulations,'' {\em arXiv preprint
  arXiv:2210.07237}, 2022.

\bibitem{P6150}
O.~Wohlfahrt, C.~Dellago, and M.~Sega, ``Ab initio structure and thermodynamics
  of the rpbe-d3 water/vapor interface by neural-network molecular dynamics,''
  {\em J. Chem. Phys.}, vol.~153, no.~14, p.~144710, 2020.

\bibitem{donkor2023machine}
E.~D. Donkor, A.~Laio, and A.~Hassanali, ``Do machine-learning atomic
  descriptors and order parameters tell the same story? the case of liquid
  water,'' {\em J. Chem. Theory Comput.}, 2023.

\bibitem{Bonn_Backus_2015}
M.~Bonn, Y.~Nagata, and E.~H. Backus, ``Molecular structure and dynamics of
  water at the water--air interface studied with surface-specific vibrational
  spectroscopy,'' {\em Angew. Chem. Int. Ed.}, vol.~54, no.~19, pp.~5560--5576,
  2015.

\bibitem{P6644}
T.~Ohto, K.~Usui, T.~Hasegawa, M.~Bonn, and Y.~Nagata, ``Toward ab initio
  molecular dynamics modeling for sum-frequency generation spectra; an
  efficient algorithm based on surface-specific velocity-velocity correlation
  function,'' {\em J. Chem. Phys.}, vol.~143, no.~12, p.~124702, 2015.

\bibitem{P6134}
S.~Shepherd, J.~Lan, D.~M. Wilkins, and V.~Kapil, ``Efficient quantum
  vibrational spectroscopy of water with high-order path integrals: From bulk
  to interfaces,'' {\em J. Phys. Chem. Lett.}, vol.~12, pp.~9108--9114, 2021.

\bibitem{P6643}
Y.~Litman, J.~Lan, Y.~Nagata, and D.~M. Wilkins, ``Fully first-principles
  surface spectroscopy with machine learning,'' {\em J. Phys. Chem. Lett.},
  vol.~14, no.~36, pp.~8175--8182, 2023.

\bibitem{P6152}
S.~P. Niblett, M.~Galib, and D.~T. Limmer, ``Learning intermolecular forces at
  liquid–vapor interfaces,'' {\em J. Chem. Phys.}, vol.~155, no.~16,
  p.~164101, 2021.

\bibitem{Sega2016}
M.~Sega and C.~Dellago, ``Long-range dispersion effects on the water/vapor
  interface simulated using the most common models,'' {\em J. Phys. Chem. B},
  vol.~121, 12 2016.

\bibitem{P6649}
I.~Sanchez-Burgos, M.~C. Muniz, J.~R. Espinosa, and A.~Z. Panagiotopoulos, ``A
  deep potential model for liquid--vapor equilibrium and cavitation rates of
  water,'' {\em J. Chem. Phys.}, vol.~158, p.~184504, 2023.

\bibitem{P6650}
M.~de~la Puente and D.~Laage, ``How the acidity of water droplets and films is
  controlled by the air--water interface,'' {\em J. Am. Chem. Soc.}, vol.~145,
  pp.~25186--25194, 2023.

\bibitem{P6007}
Y.~Shao, L.~Knijff, F.~M. Dietrich, K.~Hermansson, and C.~Zhang, ``Modelling
  bulk electrolytes and electrolyte interfaces with atomistic machine
  learning,'' {\em Batter. Supercaps}, vol.~4, pp.~1--12, 2021.

\bibitem{P4670}
M.~Hellström and J.~Behler, ``Concentration-dependent proton transfer
  mechanisms in aqueous naoh solutions: From acceptor-driven to donor-driven
  and back,'' {\em J. Phys. Chem. Lett.}, vol.~7, pp.~3302--3306, 2016.

\bibitem{P4895}
M.~Hellström and J.~Behler, ``Structure of aqueous naoh solutions: insights
  from neural-network-based molecular dynamics simulations,'' {\em Phys. Chem.
  Chem. Phys.}, vol.~19, p.~82, 2017.

\bibitem{P5126}
M.~Hellström and J.~Behler, ``Proton-transfer-driven water exchange mechanism
  in the na+ solvation shell,'' {\em J. Phys. Chem. B}, vol.~121, p.~4184,
  2017.

\bibitem{P5631}
M.~Hellström, M.~Ceriotti, and J.~Behler, ``Nuclear quantum effects in sodium
  hydroxide solutions from neural network molecular dynamics simulations,''
  {\em J. Phys. Chem. B}, vol.~122, pp.~10158--10171, 2018.

\bibitem{P6138}
Y.~Shao, M.~Hellström, A.~Yllö, J.~Mindemark, K.~Hermansson, J.~Behler, and
  C.~Zhang, ``Temperature effects on the ionic conductivity in concentrated
  alkaline electrolyte solutions,'' {\em Phys. Chem. Chem. Phys.}, vol.~22,
  p.~10426, 2020.

\bibitem{o2022crumbling}
N.~O'Neill, C.~Schran, S.~J. Cox, and A.~Michaelides, ``Crumbling crystals: On
  the dissolution mechanism of nacl in water,'' {\em arXiv preprint
  arXiv:2211.04345}, 2022.

\bibitem{P5615}
M.~Xu, T.~Zhu, and J.~Z. Zhang, ``Molecular dynamics simulation of zinc ion in
  water with an ab initio based neural network potential,'' {\em J. Phys. Chem.
  A}, vol.~123, pp.~6587--6595, 2019.

\bibitem{wang2023structures}
P.~Wang, Y.~Su, R.~Shi, X.~Huang, and J.~Zhao, ``Structures and spectroscopic
  properties of hydrated zinc (ii) ion clusters [zn2+ (h2o) n (n= 1- 8)] by ab
  initio study,'' {\em J. Clust. Sci.}, vol.~34, no.~3, pp.~1625--1632, 2023.

\bibitem{P3582}
E.~Kamarchik, Y.~Wang, and J.~M. Bowman, ``Quantum vibrational analysis and
  infrared spectra of microhydrated sodium ions using an ab initio potential,''
  {\em J. Chem. Phys.}, vol.~134, p.~114311, 2011.

\bibitem{P6292}
C.~Zhang, S.~Yue, A.~Z. Panagiotopoulos, M.~L. Klein, and X.~Wu, ``Dissolving
  salt is not equivalent to applying a pressure on water,'' {\em Nat. Commun.},
  vol.~13, p.~822, 2022.

\bibitem{zhang2023dissolving}
C.~Zhang, S.~Yue, A.~Z. Panagiotopoulos, M.~L. Klein, and X.~Wu, ``Why
  dissolving salt in water decreases its dielectric permittivity,'' {\em Phys.
  Rev. Lett.}, vol.~131, no.~7, p.~076801, 2023.

\bibitem{galib2021reactive}
M.~Galib and D.~T. Limmer, ``Reactive uptake of n2o5 by atmospheric aerosol is
  dominated by interfacial processes,'' {\em Science}, vol.~371, no.~6532,
  pp.~921--925, 2021.

\bibitem{avula2023understanding}
N.~V. Avula, M.~L. Klein, and S.~Balasubramanian, ``Understanding the anomalous
  diffusion of water in aqueous electrolytes using machine learned
  potentials,'' {\em J. Phys. Chem. Lett.}, vol.~14, pp.~9500--9507, 2023.

\bibitem{P4028}
N.~Artrith and A.~M. Kolpak, ``Understanding the composition and activity of
  electrocatalytic nanoalloys in aqueous solvents: A combination of dft and
  accurate neural network potentials,'' {\em Nano Lett.}, vol.~14,
  pp.~2670--2676, 2014.

\bibitem{P4859}
S.~K. Natarajan and J.~Behler, ``Neural network molecular dynamics simulations
  of solid-liquid interfaces: water at low-index copper surfaces,'' {\em Phys.
  Chem. Chem. Phys.}, vol.~18, p.~28704, 2016.

\bibitem{P4886}
S.~Kondati~Natarajan and J.~Behler, ``Self-diffusion of surface defects at
  copper-water interfaces,'' {\em J. Phys. Chem. C}, vol.~121, p.~4368, 2017.

\bibitem{P4988}
V.~Quaranta, M.~Hellström, and J.~Behler, ``Proton transfer mechanisms at the
  water-zno interface: The role of presolvation,'' {\em J. Phys. Chem. Lett.},
  vol.~8, p.~1476, 2017.

\bibitem{P5601}
V.~Quaranta, J.~Behler, and M.~Hellström, ``Structure and dynamics of the
  liquid-water/zinc-oxide interface from machine learning potential
  simulations,'' {\em J. Phys. Chem. C}, vol.~123, p.~1293, 2019.

\bibitem{P6271}
M.~F. Calegari~Andrade, H.-Y. Ko, L.~Zhang, R.~Car, and A.~Selloni, ``Free
  energy of proton transfer at the water–tio2 interface from ab initio deep
  potential molecular dynamics,'' {\em Chem. Sci.}, vol.~11, pp.~2335--2341,
  2020.

\bibitem{P6473}
B.~Wen, M.~F.~C. Andrade, L.-M. Liu, and A.~Selloni, ``Water dissociation at
  the water–rutile tio2(110) interface from ab initio-based deep neural
  network simulations,'' {\em Proc. Natl. Acad. Sci. U. S. A.}, vol.~120,
  p.~e2212250120, 2023.

\bibitem{P6481}
Y.-B. Zhuang, R.-H. Bi, and J.~Cheng, ``Resolving the odd-even oscillation of
  water dissociation at rutile tio2(110)-water interface by machine learning
  accelerated molecular dynamics,'' {\em J. Chem. Phys.}, vol.~157, p.~164701,
  2022.

\bibitem{P5600}
V.~Quaranta, M.~Hellström, J.~Behler, J.~Kullgren, P.~Mitev, and
  K.~Hermansson, ``Maximally resolved anharmonic oh vibrational spectrum of the
  water/zno(10-10) interface from a high-dimensional neural network
  potential,'' {\em J. Chem. Phys.}, vol.~148, p.~241720, 2018.

\bibitem{P5599}
M.~Hellström, V.~Quaranta, and J.~Behler, ``One-dimensional vs.
  two-dimensional proton transport processes at solid-liquid zinc-oxide-water
  interfaces,'' {\em Chem. Sci.}, vol.~10, p.~1232, 2019.

\bibitem{P6141}
M.~Eckhoff and J.~Behler, ``Insights into lithium manganese oxide-water
  interfaces using machine learning potentials,'' {\em J. Chem. Phys.},
  vol.~155, p.~244703, 2021.

\bibitem{nakanishi2023structural}
A.~Nakanishi, S.~Kasamatsu, J.~Haruyama, and O.~Sugino, ``Structural analysis
  of zirconium oxynitride/water interface using neural network potential,''
  {\em arXiv preprint arXiv:2307.11296}, 2023.

\bibitem{o2023elucidating}
C.~R. O’Connor, M.~F. Calegari~Andrade, A.~Selloni, and G.~A. Kimmel,
  ``Elucidating the water--anatase tio2 (101) interface structure using
  infrared signatures and molecular dynamics,'' {\em J. Chem. Phys.}, vol.~159,
  no.~10, 2023.

\bibitem{Li2023thermal}
Z.~Li, J.~Wang, C.~Yang, L.~Liu, and J.-Y. Yang, ``{Thermal transport across
  TiO2–H2O interface involving water dissociation: Ab initio-assisted deep
  potential molecular dynamics},'' {\em J. Chem. Phys.}, vol.~159, p.~144701,
  10 2023.

\bibitem{zeng2023mechanistic}
Z.~Zeng, F.~Wodaczek, K.~Liu, F.~Stein, J.~Hutter, J.~Chen, and B.~Cheng,
  ``Mechanistic insight on water dissociation on pristine low-index tio2
  surfaces from machine learning molecular dynamics simulations,'' {\em Nat.
  Commun.}, vol.~14, no.~1, p.~6131, 2023.

\bibitem{ding2023modeling}
Z.~Ding and A.~Selloni, ``Modeling the aqueous interface of amorphous tio2
  using deep potential molecular dynamics,'' {\em J. Chem. Phys.}, vol.~159,
  no.~2, 2023.

\bibitem{P6474}
A.~E.~G. Mikkelsen, J.~Schiøtz, T.~Vegge, and K.~W. Jacobsen, ``Is the
  water/pt(111) interface ordered at room temperature?,'' {\em J. Chem. Phys.},
  vol.~155, p.~224701, 2021.

\bibitem{P6469}
A.~E.~G. Mikkelsen, H.~H. Kristoffersen, J.~Schiøtz, T.~Vegge, H.~A. Hansen,
  and K.~W. Jacobsen, ``Structure and energetics of liquid water-hydroxyl
  layers on pt(111),'' {\em Phys. Chem. Chem. Phys.}, vol.~24, pp.~9885--9890,
  2022.

\bibitem{schienbein2022nanosecond}
P.~Schienbein and J.~Blumberger, ``Nanosecond solvation dynamics of the
  hematite/liquid water interface at hybrid dft accuracy using committee neural
  network potentials,'' {\em Phys. Chem. Chem. Phys.}, vol.~24, no.~25,
  pp.~15365--15375, 2022.

\bibitem{li2023characterizing}
L.~Li, M.~F. Calegari~Andrade, R.~Car, A.~Selloni, and E.~A. Carter,
  ``Characterizing structure-dependent tis2/water interfaces using
  deep-neural-network-assisted molecular dynamics,'' {\em J. Phys. Chem. C},
  2023.

\bibitem{raman2023acid}
A.~S. Raman and A.~Selloni, ``Acid--base chemistry of a model iro2 catalytic
  interface,'' {\em J. Phys. Chem. Lett.}, vol.~14, pp.~7787--7794, 2023.

\bibitem{P6518}
X.-T. Fan, X.-J. Wen, Y.-B. Zhuang, and J.~Cheng, ``Molecular insight into the
  gap(110)-water interface using machine learning accelerated molecular
  dynamics,'' {\em J. Energy Chem.}, vol.~in press, 2023.

\bibitem{piaggi2023first}
P.~M. Piaggi, A.~Selloni, A.~Panagiotopoulos, R.~Car, and P.~G. Debenedetti,
  ``A first-principles machine-learning force field for heterogeneous ice
  nucleation on microcline feldspar,'' {\em Faraday Discuss.}, 2023.

\bibitem{Li2023thermal.Cu}
Z.~Li, X.~Tan, Z.~Fu, L.~Liu, and J.-Y. Yang, ``Thermal transport across
  copper--water interfaces according to deep potential molecular dynamics,''
  {\em Phys. Chem. Chem. Phys.}, vol.~25, no.~9, pp.~6746--6756, 2023.

\bibitem{P6470}
X.~Yang, A.~Bhowmik, T.~Vegge, and H.~A. Hansen, ``Neural network potentials
  for accelerated metadynamics of oxygen reduction kinetics at au--water
  interfaces,'' {\em Chem. Sci.}, vol.~14, no.~14, pp.~3913--3922, 2023.

\bibitem{li2020machine}
X.~Li, W.~Paier, and J.~Paier, ``Machine learning in computational surface
  science and catalysis: Case studies on water and metal--oxide interfaces,''
  {\em Front. Chem.}, vol.~8, p.~601029, 2020.

\bibitem{P5876}
H.~Ghorbanfekr, J.~Behler, and F.~M. Peeters, ``Insights into water permeation
  through hbn nanocapillaries by ab initio machine learning molecular dynamics
  simulations,'' {\em J. Phys. Chem. Lett.}, vol.~11, p.~7363, 2020.

\bibitem{P6327}
W.~Zhao, H.~Qiu, and W.~Guo, ``A deep neural network potential for water
  confined in graphene nanocapillaries,'' {\em J. Phys. Chem. C}, 2022.

\bibitem{liu2023transferability}
D.~Liu, J.~Wu, and D.~Lu, ``Transferability evaluation of the deep potential
  model for simulating water-graphene confined system,'' {\em J. Chem. Phys.},
  vol.~159, no.~4, 2023.

\bibitem{kapil2022first}
V.~Kapil, C.~Schran, A.~Zen, J.~Chen, C.~J. Pickard, and A.~Michaelides, ``The
  first-principles phase diagram of monolayer nanoconfined water,'' {\em
  Nature}, vol.~609, no.~7927, pp.~512--516, 2022.

\bibitem{P6464}
F.~L. Thiemann, C.~Schran, P.~Rowe, E.~A. Müller, and A.~Michaelides, ``Water
  flow in single-wall nanotubes: Oxygen makes it slip, hydrogen makes it
  stick,'' {\em ACS Nano}, vol.~16, pp.~10775--10782, 2022.

\bibitem{cao2023neural}
Z.~Cao, Y.~Wang, C.~Lorsung, and A.~Barati~Farimani, ``Neural network predicts
  ion concentration profiles under nanoconfinement,'' {\em J. Chem. Phys.},
  vol.~159, no.~9, 2023.

\bibitem{P6145}
L.~Shen and W.~Yang, ``Molecular dynamics simulations with quantum
  mechanics/molecular mechanics and adaptive neural networks,'' {\em J. Chem.
  Theor. Comp.}, vol.~14, pp.~1442--1455, 2018.

\bibitem{P6288}
M.~Yang, L.~Bonati, D.~Polino, and M.~Parrinello, ``Using metadynamics to build
  neural network potentials for reactive events: the case of urea decomposition
  in water,'' {\em Catal. Today}, vol.~387, pp.~143--149, 2022.

\bibitem{P6573}
T.~Devergne, T.~Magrino, F.~Pietrucci, and A.~M. Saitta, ``Combining machine
  learning approaches and accurate ab initio enhanced sampling methods for
  prebiotic chemical reactions in solution,'' {\em J. Chem. Theor. Comp.},
  vol.~18, pp.~5410--5421, 2022.

\bibitem{P6297}
X.~Pan, J.~Yang, R.~Van, E.~Epifanovsky, J.~Ho, J.~Huang, J.~Pu, Y.~Mei,
  K.~Nam, and Y.~Shao, ``Machine-learning-assisted free energy simulation of
  solution-phase and enzyme reactions,'' {\em J. Chem. Theor. Comp.}, vol.~17,
  pp.~5745--5758, 2021.

\bibitem{lan2021simulating}
J.~Lan, V.~Kapil, P.~Gasparotto, M.~Ceriotti, M.~Iannuzzi, and V.~V. Rybkin,
  ``Simulating the ghost: quantum dynamics of the solvated electron,'' {\em
  Nat. Commun.}, vol.~12, no.~1, p.~766, 2021.

\bibitem{chen2019integrating}
W.-K. Chen, W.-H. Fang, and G.~Cui, ``Integrating machine learning with the
  multilayer energy-based fragment method for excited states of large
  systems,'' {\em J. Phys. Chem. Lett.}, vol.~10, no.~24, pp.~7836--7841, 2019.

\bibitem{wang2019machine}
J.~Wang, S.~Olsson, C.~Wehmeyer, A.~P{\'e}rez, N.~E. Charron, G.~De~Fabritiis,
  F.~No{\'e}, and C.~Clementi, ``Machine learning of coarse-grained molecular
  dynamics force fields,'' {\em ACS Cent. Sci.}, vol.~5, no.~5, pp.~755--767,
  2019.

\bibitem{husic2020coarse}
B.~E. Husic, N.~E. Charron, D.~Lemm, J.~Wang, A.~P{\'e}rez, M.~Majewski,
  A.~Kr{ä}mer, Y.~Chen, S.~Olsson, G.~de~Fabritiis, {\em et~al.}, ``Coarse
  graining molecular dynamics with graph neural networks,'' {\em J. Chem.
  Phys.}, vol.~153, no.~19, 2020.

\bibitem{krämer2023statistically}
A.~krämer, A.~E. Durumeric, N.~E. Charron, Y.~Chen, C.~Clementi, and
  F.~No{\'e}, ``Statistically optimal force aggregation for coarse-graining
  molecular dynamics,'' {\em J. Phys. Chem. Lett.}, vol.~14, no.~17,
  pp.~3970--3979, 2023.

\bibitem{yao2023machine}
S.~Yao, R.~Van, X.~Pan, J.~H. Park, Y.~Mao, J.~Pu, Y.~Mei, and Y.~Shao,
  ``Machine learning based implicit solvent model for aqueous-solution alanine
  dipeptide molecular dynamics simulations,'' {\em RSC Adv.}, vol.~13, no.~7,
  pp.~4565--4577, 2023.

\bibitem{P6133}
J.~R. Cendagorta, H.~Shen, Z.~Bacic, and M.~E. Tuckerman, ``Enhanced sampling
  path integral methods using neural network potential energy surfaces with
  application to diffusion in hydrogen hydrates,'' {\em Adv. Theory Sim.},
  vol.~4, p.~2000258, 2021.

\bibitem{yang2023machine}
J.~Yang, Y.~Cong, Y.~Li, and H.~Li, ``Machine learning approach based on a
  range-corrected deep potential model for efficient vibrational frequency
  computation,'' {\em J. Chem. Theory Comput.}, vol.~19, no.~18,
  pp.~6366--6374, 2023.

\bibitem{boselt2021machine}
L.~Böselt, M.~Thürlemann, and S.~Riniker, ``Machine learning in qm/mm
  molecular dynamics simulations of condensed-phase systems,'' {\em J. Chem.
  Theory Comput.}, vol.~17, no.~5, pp.~2641--2658, 2021.

\bibitem{hofstetter2022graph}
A.~Hofstetter, L.~B{ö}selt, and S.~Riniker, ``Graph-convolutional neural
  networks for (qm) ml/mm molecular dynamics simulations,'' {\em Phys. Chem.
  Chem. Phys.}, vol.~24, no.~37, pp.~22497--22512, 2022.

\bibitem{salehi2022hydration}
S.~M. Salehi, S.~K{ä}ser, K.~T{ö}pfer, P.~Diamantis, R.~Pfister, P.~Hamm,
  U.~Rothlisberger, and M.~Meuwly, ``Hydration dynamics and ir spectroscopy of
  4-fluorophenol,'' {\em Phys. Chem. Chem. Phys.}, vol.~24, no.~42,
  pp.~26046--26060, 2022.

\bibitem{xu2021automatically}
M.~Xu, T.~Zhu, and J.~Z. Zhang, ``Automatically constructed neural network
  potentials for molecular dynamics simulation of zinc proteins,'' {\em Front.
  Chem.}, vol.~9, p.~692200, 2021.

\bibitem{loeffler2021conformational}
J.~R. Loeffler, M.~L. Fern{\'a}ndez-Quintero, F.~Waibl, P.~K. Quoika, F.~Hofer,
  M.~Schauperl, and K.~R. Liedl, ``Conformational shifts of stacked
  heteroaromatics: vacuum vs. water studied by machine learning,'' {\em Front.
  Chem.}, vol.~9, p.~641610, 2021.

\bibitem{gao2021graphical}
P.~Gao, X.~Yang, Y.-H. Tang, M.~Zheng, A.~Andersen, V.~Murugesan, A.~Hollas,
  and W.~Wang, ``Graphical gaussian process regression model for aqueous
  solvation free energy prediction of organic molecules in redox flow
  batteries,'' {\em Phys. Chem. Chem. Phys.}, vol.~23, no.~43,
  pp.~24892--24904, 2021.

\bibitem{P6074}
A.~Fabrizio, A.~Grisafi, B.~Meyer, M.~Ceriotti, and C.~Corminboeuf, ``Electron
  density learning of non-covalent systems,'' {\em Chem. Sci.}, vol.~10,
  pp.~9424--9432, 2019.

\bibitem{P5629}
A.~Grisafi and M.~Ceriotti, ``Incorporating long-range physics in atomic-scale
  machine learning,'' {\em J. Chem. Phys.}, vol.~151, p.~204105, 2019.

\bibitem{P6542}
T.~W. Ko, J.~A. Finkler, S.~Goedecker, and J.~Behler, ``Accurate
  fourth-generation machine learning potentials by electrostatic embedding,''
  {\em J. Chem. Theory Comput.}, vol.~19, pp.~3567--3579, 2023.

\bibitem{ARPC_Noe_2020}
F.~No\'{e}, A.~Tkatchenko, K.-R. M\"{u}ller, and C.~Clementi, ``Machine
  learning for molecular simulation,'' {\em Annu. Rev. Phys. Chem.}, vol.~71,
  no.~1, pp.~361--390, 2020.
\newblock PMID: 32092281.

\bibitem{JCP_Geiger_2013}
P.~Geiger and C.~Dellago, ``{Neural networks for local structure detection in
  polymorphic systems},'' {\em J. Chem. Phys.}, vol.~139, p.~164105, 10 2013.

\bibitem{JCIM_Fulford_2019}
M.~Fulford, M.~Salvalaglio, and C.~Molteni, ``Deepice: A deep neural network
  approach to identify ice and water molecules,'' {\em J. Chem. Inf. Model.},
  vol.~59, no.~5, pp.~2141--2149, 2019.
\newblock PMID: 30875217.

\bibitem{noe2019boltzmann}
F.~No{\'e}, S.~Olsson, J.~K{ö}hler, and H.~Wu, ``Boltzmann generators:
  Sampling equilibrium states of many-body systems with deep learning,'' {\em
  Science}, vol.~365, no.~6457, p.~eaaw1147, 2019.

\bibitem{MLST_Wirnsberger_2022}
P.~Wirnsberger, G.~Papamakarios, B.~Ibarz, S.~Racanière, A.~J. Ballard,
  A.~Pritzel, and C.~Blundell, ``Normalizing flows for atomic solids,'' {\em
  Mach. learn.: sci. technol.}, vol.~3, p.~025009, may 2022.

\bibitem{zeni2023mattergen}
C.~Zeni, R.~Pinsler, D.~Z{ü}gner, A.~Fowler, M.~Horton, X.~Fu, S.~Shysheya,
  J.~Crabb{\'e}, L.~Sun, J.~Smith, {\em et~al.}, ``Mattergen: a generative
  model for inorganic materials design,'' {\em arXiv preprint
  arXiv:2312.03687}, 2023.

\bibitem{PNAS_Tiwary_2023}
Z.~Zou, E.~R. Beyerle, S.-T. Tsai, and P.~Tiwary, ``Driving and characterizing
  nucleation of urea and glycine polymorphs in water,'' {\em Proc. Natl. Acad.
  Sci. U.S.A.}, vol.~120, no.~7, p.~e2216099120, 2023.

\bibitem{NatCompSci_Jung_2023}
H.~Jung, R.~Covino, A.~Arjun, C.~Leitold, C.~Dellago, P.~G. Bolhuis, and
  G.~Hummer, ``Machine-guided path sampling to discover mechanisms of molecular
  self-organization,'' {\em Nat. Comput. Sci.}, vol.~3, no.~4, pp.~334--345,
  2023.

\end{thebibliography}
\bibliographystyle{ieeetr} 

\end{document}